\documentclass[12pt]{article}

\usepackage{psfig}
\usepackage{graphicx}
\usepackage{deluxetable1}

\usepackage{color}

\usepackage[normalem]{ulem}
\usepackage{amsthm,amsmath,amssymb}

\def\lsim{\hbox{ \rlap{\raise 0.425ex\hbox{$<$}}\lower 0.65ex\hbox{$\sim$} }}
\def\gsim{\hbox{ \rlap{\raise 0.425ex\hbox{$>$}}\lower 0.65ex\hbox{$\sim$} }}

\def\arcsec{\hbox{$^{\prime\prime}$}}

\def\f(h{\hbox{$~\!\!^{\rm h}$}}

\def\ale{\mathrel{\hbox{\rlap{\hbox{\lower4pt\hbox{$\sim$}}}\hbox{$<$}}}}
\def\age{\mathrel{\hbox{\rlap{\hbox{\lower4pt\hbox{$\sim$}}}\hbox{$>$}}}}

\usepackage{multicol}
\usepackage{aas_long,psfig}
\usepackage[numbers,sort&compress]{natbib}
\def\gtrsim{\mathrel{\hbox{\rlap{\hbox{\lower4pt\hbox{$\sim$}}}\hbox{$>$}}}}
\def\lessim{\mathrel{\hbox{\rlap{\hbox{\lower4pt\hbox{$\sim$}}}\hbox{$<$}}}}

		\usepackage{psfig}

		\usepackage{color,calc}
		  \definecolor{myColor}{rgb}{0.9,0.9,0.9}%   rgb color model
		   {\end{minipage}%
		   \end{lrbox}%
		   \colorbox{myColor}{\usebox{\@tempboxa}}}%

		\definecolor{Light}{gray}{.80}
		
\usepackage{times}
\usepackage[light,bottom,none]{draftcopy}
\usepackage{epsfig, amssymb, amsmath, multicol}
\usepackage{color}
\usepackage{wrapfig} 
\usepackage{hyperref}

\newcommand{\mmsun}{M_{\odot}}

\def\h2{H$_2$}
\def\f0{$F_0$}

\newcommand{\sci}[1]{{\rm \; \times \; 10^{#1}}}

\def\mzreion{z_{\rm reion}}
\def\sqdeg{sq.~deg}
\def\aap{A \& A}
\def\aj{AJ}
\def\apj{ApJ}
\def\apjl{ApJL}
\def\apjs{ApJS}

\def\araa{ARAA}
\def\mnras{MNRAS}
\def\nat{Nature}

\long\def\@makecaption#1#2{\vskip 2ex\noindent#1\ #2\par}% 

\def\@xfigcaption[#1]#2{{\def\@captype{figure}\caption{#2}}}% 

\usepackage{geometry}
\usepackage[layout=modern]{advancedcoverpage}

\usepackage{fancyheadings}

\vspace{-1in}
\title{\includegraphics[width=6.5in]{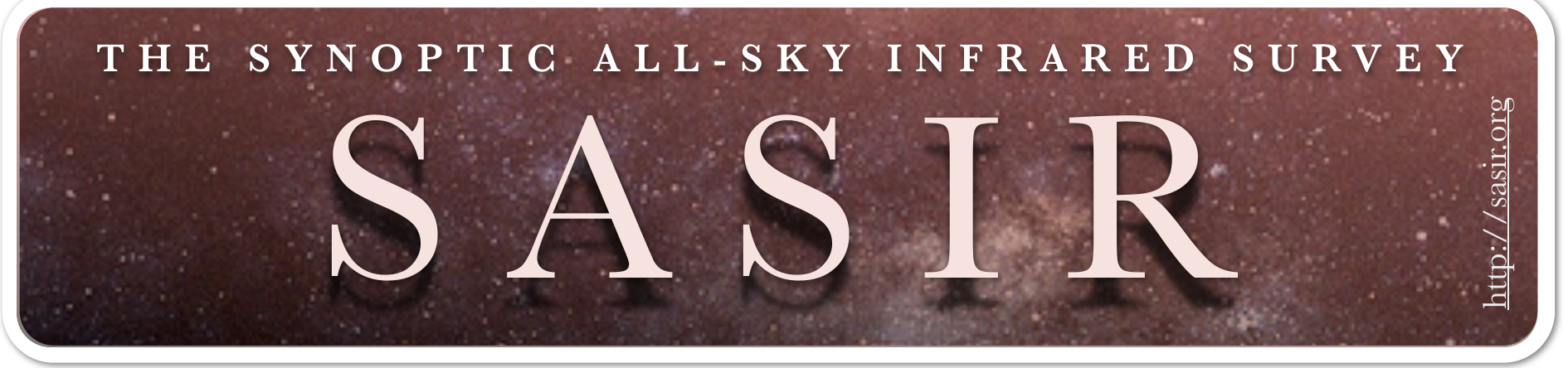}}
\subtitle{A Collaboration Between the University of California, the Mexican Astronomy Community, and the University of Arizona.\\  Submitted to the ``Optical and IR Astronomy from the
  Ground'' Program Prioritization Panel of the Astro2010 Decadal
  Survey}
\author{\emph{\underline{Authors}}\\ Joshua S.\ Bloom
  {\small (Department of Astronomy, UC Berkeley; \href{mailto:jbloom@astro.berkeley.edu}{jbloom@astro.berkeley.edu};
  +1-510-643-4621)}, \\ J. Xavier Prochaska (UCO/Lick), William Lee
  (IA-UNAM), Jesus Gonz\'alez (IA-UNAM), Enrico Ram\'irez-Ruiz (UC Santa
  Cruz), Michael Bolte (UC Santa Cruz/UCO/Lick), Jos\'e Franco
  (IA-UNAM), Jos\'e Guichard (INAOE), Alberto Carrami\~nana (INAOE),
  Peter Strittmatter (U.\ Arizona), 
  \\ Vladimir Avila-Reese (IA-UNAM), Rebecca Bernstein (UCO/Lick),
  Bruce Bigelow (UCO/Lick), Mark Brodwin (CfA), Adam Burgasser (UC San
  Diego), Nat Butler (UC Berkeley), Miguel Ch\'avez (INAOE), Bethany Cobb
  (UC Berkeley), Kem Cook (LLNL), Irene Cruz-Gonz\'alez (IA-UNAM),
  Jos\'e Antonio de Diego (IA-UNAM), Alejandro Farah (IA-UNAM), Leonid
  Georgiev (IA-UNAM), Julien Girard (ESFM-IPN), Hector
  Hern\'andez-Toledo (IA-UNAM), Elena Jim\'enez-Bail\'on (IA-UNAM), Yair
  Krongold, (IA-UNAM), Divakara Mayya (INAOE), Juan Meza (LBNL),
  Takamitsu Miyaji (IA-UNAM), Ra\'ul M\'ujica (INAOE), Peter Nugent
  (LBNL), Alicia Porras (INAOE), Dovi Poznanski (UC Berkeley/LBNL),
  Alejandro Raga (ICN-UNAM), Michael Richer (IA-UNAM), Lino Rodr\'iguez
  (INAOE), Daniel Rosa (INAOE), Adam Stanford (UC Davis), Andrew
  Szentgyorgyi (CfA), Guillermo Tenorio-Tagle (INAOE), Rollin Thomas
  (LBNL), Octavio Valenzuela (IA-UNAM), Alan Watson (CRyAUNAM, IA-UNAM), Peter Wehinger (U.\ Arizona)}

\date{}

\usepackage{makeidx} 
\usepackage[cols=3]{glossary}
\makeindex
\makeglossary
\usepackage{setspace}

\begin{document}
\maketitle

\def\ale{\mathrel{\hbox{\rlap{\hbox{\lower4pt\hbox{$\sim$}}}\hbox{$<$}}}}
\def\age{\mathrel{\hbox{\rlap{\hbox{\lower4pt\hbox{$\sim$}}}\hbox{$>$}}}}

\newcommand\ion[2]{#1$\;${\small\rmfamily\sc{#2}}\relax}% 

\newcommand{\avgchi}{\bar{\chi}_{\rm H I}}
\newcommand{\barxi}{{\bar{x}_i}}
\newcommand{\Mpc}{{\rm Mpc}}
\newcommand{\MHz}{{\rm MHz}}
\newcommand{\Msun}{{M_{\odot}}}
\newcommand{\bfM}{{\boldsymbol{M}}}
\newcommand{\bfP}{{\boldsymbol{P}}}
\def\apj{ApJ}
\def\apjl{ApJL}
\def\apjs{ApJS}
\def\aj{AJ}
\def\mnras{MNRAS}
\def\physrep{physrep}
\def\pasj{PASJ}
\def\araa{{Ann.\ Rev.\ Astron.\& Astrophys.\ }}
\def\aap{{\em A.\&A}}
\def\prd{PRD}

\vspace{-0.3cm}
\thispagestyle{empty}

\textwidth 6.5in 
\textheight 9.00in 
\voffset = -0.90in 
\headsep = 0.2in 
\headheight=0.2in
\topmargin=0.5in

\section{Summary}
\vspace{-0.12in}
We are proposing to conduct a simultaneous multicolor ($Y$, $J$, $H$,
$K$) synoptic infrared (IR)\glossary{name={IR},description={Infrared}}
imaging survey of the entire sky (above declination $\delta =
-30^\circ$) with a new, dedicated 6.5-meter telescope at San Pedro
M\'artir (SPM)\index{San Pedro M\'artir
Observatory}\glossary{name={SPM},description={San Pedro M\'artir}}
Observatory (Mexico).  This initiative is being developed in
partnership with astronomy institutions in Mexico,  the University
of California, and the University of Arizona.  This 4--5 year, dedicated survey, planned to begin in
2017, will reach more than 100 times deeper than
2MASS\index{2MASS}\glossary{name={2MASS},description={Two-Micron
All-Sky Survey}} \cite{scs+06}, increasing the effective detection
volume by more than one million. {The Synoptic All-Sky Infrared
(SASIR) Survey will reveal the missing sample of faint red dwarf stars
in the local solar neighborhood, and the unprecedented sensitivity
over such a wide field will result in the discovery of thousands of
$z \sim 7$ quasars (and reaching to $z>10$), allowing detailed study
(in concert with
JWST\index{JWST}\glossary{name={JWST},description={James Webb Space
Telescope}} and Giant Segmented Mirror Telescopes\index{Giant
Segmented Mirror Telescopes}) of the timing and the origin(s) of
reionization in the early universe. As a time-domain survey, SASIR
will reveal the dynamic infrared universe as never seen before,
opening new phase space for discovery. Moreover, synoptic observations
of over 10$^6$ supernovae and variable stars will provide better
distance measures than optical studies alone.

SASIR also provides significant synergy with other major Astro2010
facilities, improving the overall scientific return of community
investments. Compared to optical-only measurements, IR colors vastly
improve photometric redshifts to $z \approx 4$, enhancing dark energy
and dark matter surveys based on weak lensing and baryon
oscillations. The wide field and ToO capabilities will enable a
connection of the gravitational wave (e.g., Advanced LIGO\index{LIGO}
and LISA\index{LISA}) and neutrino universe -- with events otherwise
poorly localized on the sky -- to transient electromagnetic
phenomena. SASIR will enable the distribution of dust to be mapped
more precisely and with higher dynamic range than currently possible,
removing systematic bias in extragalactic distance and galaxy studies.

\smallskip
\noindent {\it \bf Technical Overview}: The 6.5m primary mirror 
is already funded and casting will reach the peak temperature stage in late August 2009.  The SASIR
telescope and dome structure will be based on the
Magellan\index{Magellan} or MMT\index{MMT} design --- with
demonstrated capability to deliver excellent image quality in an f/5
beam over 1+ degree diameter.  This will mitigate risk and speed
development.  The camera consists of reimaging optics, 3
dichroics and 4 separate focal planes each seeing $\sim1$ degree
diameter of the sky with an effective focal ratio of f/2.5. In total,
there will be 124 2k $\times$ 2k arrays, constituting the largest IR
imager ever constructed. Thermal emission control and weight/space
constraints will present significant engineering challenges, but no
new technical development is required. The \hbox{{\it \'etendue-couleur}} is
more than 3 orders of magnitude larger than 2MASS and $>$10$\times$ that
of VISTA. Still, data rates ($\sim$TB/night) are roughly 100$\times$
smaller than those expected from LSST and can be accommodated with
ongoing upgrades at SPM. Data management and archiving will be
performed in the US. We expect to release survey data to the US and
Mexican astronomy communities in incremental stages throughout the
science operations, and to release transients at least as often as
daily.

\smallskip
\noindent {\it \bf Current State of funding}: 
SASIR is ``pre-phase A,'' having finished a {\it preliminary
conceptual design} phase.  The primary mirror is fully funded, but no
other significant funding has been secured for this project to date.
Support for the conceptual and preliminary design phases was solicited (in late 2008)
from the NSF\index{NSF} and the Consejo Nacional de Ciencia y
Tecnologia (CONACyT\index{CONACyT}). Funding for a pathfinder
multi-color optical-IR survey
(``RATIR''\glossary{name={RATIR},description={Reionization and
Transients Infrared Project}}) on the 1.5m telescope at SPM has been
secured (through grants from NASA) and first light is scheduled in early 2010.

\newpage
\section{Key Science Goals}
\pagestyle{fancy} 
\label{sec:sci}
\setcounter{page}{1}
\subsection{Unveiling the Lowest-Temperature Neighbors}
\label{sec:lm}

Our understanding of stellar populations stems from our sampling of
the immediate Solar Neighborhood. This sample is woefully incomplete for brown dwarfs
(BDs)\glossary{name={BD},description={Brown Dwarf}}, objects 
which are incapable of sustaining core Hydrogen fusion.  With masses extending \glossary{name={WP},description={Astro2010 Whitepaper}}
from $\sim 0.075 \mmsun$ to below $\sim 0.013 \mmsun$ (the hydrogen
and deuterium-burning minimum masses) BDs probe the low-mass limits
of star formation processes and serve as a bridge between stellar astrophysics and planetary science.  Lacking nuclear energy generation, they evolve steadily to
low luminosities and low effective temperatures and are thus useful
chronometers for a variety of Galactic studies (see following Astro2010 Whitepaper [WP]: \href{http://www8.nationalacademies.org/astro2010/DetailFileDisplay.aspx?id=343}{Burgasser}).  They ultimately
achieve photospheric conditions similar to those of gas giant
planets ($T_{\rm eff} \approx 100-1000$K).
Overall, studies of BD populations and their atmospheres support a wide range
of scientific endeavors: providing discriminating constraints on star-
and planet-formation theories; driving advances in the properties of
cool atmospheres; and guiding direct detection strategies for
exoplanets.   Testing models of BD atmospheres, and identifying
true analogues to directly detected exoplanets 
(e.g., Formalhaut b\index{Formalhaut b},
$T_{\rm eff} \approx 500$K; \cite{2008Sci...322.1345K})
will require detailed studies of BDs cooler than those
currently known ($T_{\rm eff} \approx 600$K).

\begin{figure}[tbh]
	\begin{minipage}{4in}
	\includegraphics[width=4in]{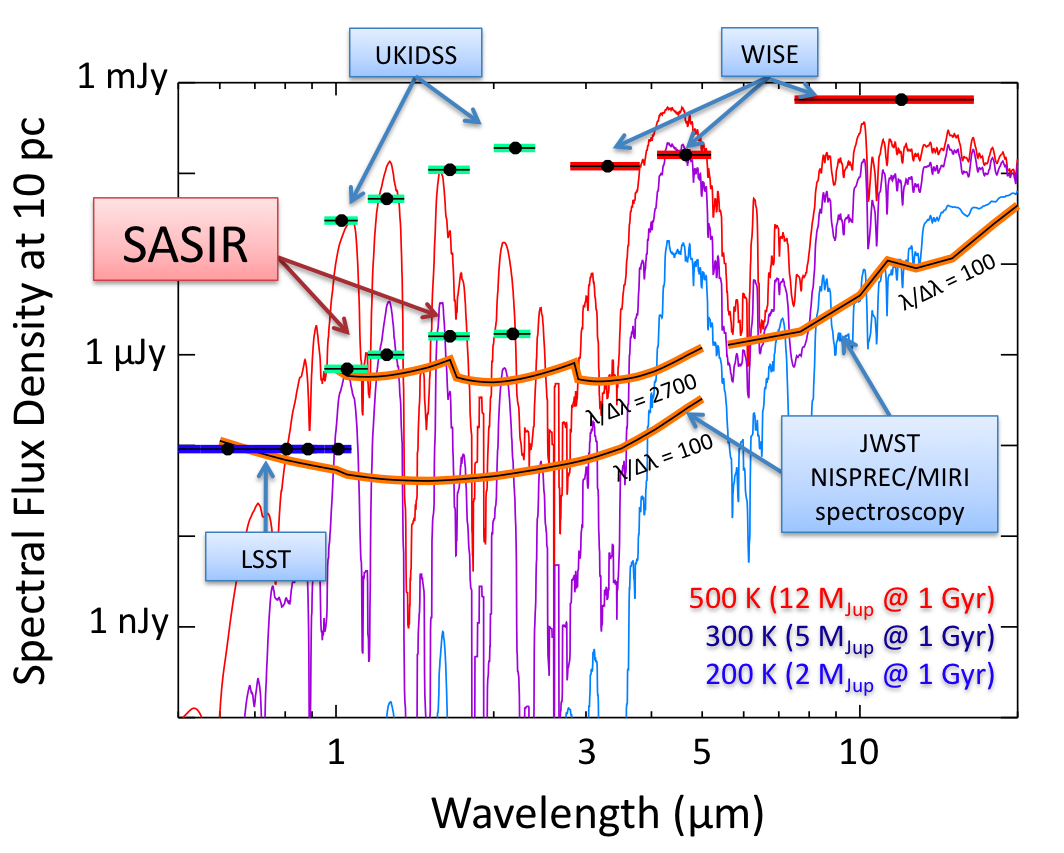}
	\end{minipage}\hspace{0.05in}
	\begin{minipage}{2.3in}
\caption{ \it
Model spectra \cite{2003ApJ...596..587B} for 500 K, 300 K and 200 K brown dwarfs 
(top to bottom), scaled 
to a distance of 10 pc.  These models correspond to masses of 
12, 5 and 2 Jupiter masses at an age of 1 Gyr, respectively.  Sensitivity 
limits for current and proposed imaging surveys (including SASIR) and spectroscopic facilities (5$\sigma$ in 1 hour) are indicated.  \label{fig:bdsrvy}
}
\end{minipage}
\end{figure}

UKIDSS\index{UKIDSS}\glossary{name={UKIDSS},description={United Kingdom Infrared Deep Sky Survey}} \cite{2007MNRAS.379.1599L} and the pending WISE\index{WISE}\glossary{name={WISE},description={Wide-Field Infrared Survey Explorer}} \cite{2004SPIE.5487..101D} experiments should detect the first
$T_{\rm eff} \approx 500$K brown dwarfs within 10\,pc
(Fig.\ ~\ref{fig:bdsrvy}), but deeper optical/NIR surveys are required
to extend samples to lower $T_{\rm eff}$ and beyond the local
neighborhood.  SASIR will provide a 100-fold increase in
  sensitivity over current NIR\glossary{name={NIR},description={Near-IR}} surveys, facilitating the detection of
1000K brown dwarfs out to 1~kpc (sampling the Galactic scale height of
BDs), 500K brown dwarfs out to 100~pc (a 1000-fold increase in sampled
volume over UKIDSS) and 300K ``water-cloud'' brown dwarfs out to
10\,pc. This will help {\bf complete the census of stars and BDs down to
  the lowest masses in the immediate Solar Neighborhood}, while
sampling the density structure of brown dwarfs far from the Galactic
plane. The multi-band and multi-epoch detections with SASIR, optimized
in cadence for parallax discovery and proper motion studies, in conjunction with
single-band detections and upper limits from WISE in the mid-IR and LSST
in the optical, will enable robust color- and motion-selection of
sources for efficient spectroscopic follow-up with GSMTs\glossary{name={GSMT},description={Giant Segmented Mirror Telescope}}\index{Giant Segmented Mirror Telescopes} and JWST\index{JWST}.  Expansion of the local census will
also enable direct imaging searches for Earth-mass exoplanets 
and other companions with next-generation high-angular resolution facilities.

\subsection{Probing the Epoch of Reionization with Quasars}

With no in situ observations, the objects responsible for the
reionization\index{Reionization} of Hydrogen beyond $z \approx 7$ span 
a wide range of
theoretical possibilities \cite{2009arXiv0902.3442M,pbf+09} (WP: \href{http://www8.nationalacademies.org/astro2010/DetailFileDisplay.aspx?id=253}{Prochaska}, \href{http://www8.nationalacademies.org/astro2010/DetailFileDisplay.aspx?id=329}{McQuinn}, \href{http://www8.nationalacademies.org/astro2010/DetailFileDisplay.aspx?id=346}{Bouwens},
\href{http://www8.nationalacademies.org/astro2010/DetailFileDisplay.aspx?id=99}{Stiavelli}).  The consensus is that bright quasars (QSOs) have insufficient number density at $z>5$ to drive reionization
\citep[e.g.,][]{2008ApJ...688...85F}, but AGN\glossary{name={AGN},description={Active Galactic Nuclei}} remain the only extragalactic source
known to have high escape fractions of ionizing radiation.  Surveys
for $z>5$ QSOs have observed the brightest sources, leaving the faint
end of the luminosity function mostly unconstrained.  If the faint end
steepens (as for high $z$ galaxies in the UV\glossary{name={UV},description={Ultraviolet}}) or if entirely different
classes of AGN \citep[e.g.\ mini-quasars;][]{2004ApJ...604..484M} exist at early
times, these would contribute to the extragalactic ultraviolet background 
at $z>5$.  Independent of reionization studies, surveys for $z>6$
QSOs are also valuable as tracers of the
growth of supermassive black holes (BHs\index{Black holes}\glossary{name={BH},description={Black hole}}) 
that populate modern galaxies or as signposts
for follow-up studies of early galaxy formation \cite{car09,bla09}.
In the next decade, it will be possible to {\bf determine the nature
and role of QSOs during reionization} with SASIR as a major contributor.

\begin{figure}[hbt]

	\begin{minipage}{2.7in}
	\caption{ \it \footnotesize Predicted surface density of
          quasars (solid black line) per 1000\,\sqdeg\ as a function
          of limiting magnitude assuming a double-power law luminosity
          function with $\beta_l = -1.64$, $\beta_h = -3.2$, with
          $M_{1450}^*=-24.5$ and $6\sci{-10} \, {\rm Mpc^{-3}}$ quasars 
brighter than
          $M=-26.7$ \citep{2004AJ....128..515F}.  The curves assume a 
          number density evolution $\propto \exp(-0.43 z)$.
          The dotted lines indicate more optimistic and pessimistic
          assumptions on the number density and redshift evolution.
          The vertical lines indicate the magnitude limits of UKIDSS\index{UKIDSS},
          VISTA\index{VISTA}, and SASIR with the number of detections given the
          proposed survey area in parenthesis. Only SASIR survey will
          have the depth and sky coverage to provide a meaningful
          sample of $z>7$ quasars and have a realistic chance of
          detecting sources at $z > 10$. 	\label{fig:highzqso}}
\end{minipage} \begin{minipage}{3.9in}
\centerline{\includegraphics[width=3.9in]{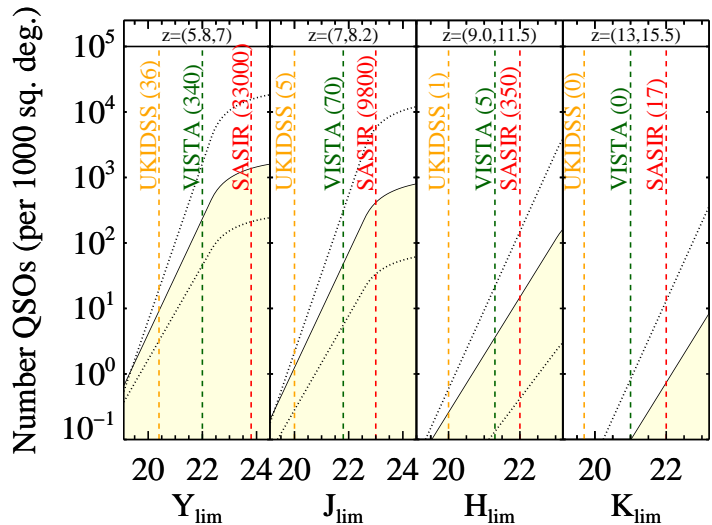}}
\end{minipage}
\end{figure}

QSO searches in the reionization epoch are challenged by three effects:
(i) cosmological dimming; (ii) low number density; and (iii) their
extremely red color.  Thus, high $z$ QSO surveys require deep near-IR
and optical imaging over a large area of sky.  
While SDSS\index{SDSS}\glossary{name={SDSS},description={Sloan Digital Sky Survey}} 
and 2DF (and the upcoming Pan-STARRS project) provide large areas of optical
imaging, no project has achieved comparable depth and area in the
near-IR.  The nearly completed UKIDSS\index{UKIDSS} and the upcoming 
VISTA\index{VISTA} \glossary{name={VISTA},description={Visible 
and Infrared Survey Telescope for Astronomy}} surveys
(a joint UK/ESO venture) should increase the current samples, but
these programs also lack sufficient depth and/or area to meaningfully
constrain the QSO\glossary{name={QSO},description={Quasi-stellar Object}} 
population, especially at $z>7$
(Fig.\ \ref{fig:highzqso}).   A systematic study of AGN during the
reionization era requires the survey characteristics of SASIR. At $z \sim
6$, such surveys would establish the QSO luminosity function (surpassing
even LSST\index{LSST} for reddened AGN) 
and are necessary for sampling even the brightest
sources at higher $z$.  These measurements would be compared against
estimates of the black-hole merger rate at $z \approx \mzreion$ from
gravitational-wave experiments (e.g.\ LISA\index{LISA}).

%%%%%%%%%%%%%%%%%%%%%%%%%%%%%%%%%%%%%%%%%%%%%%%%%%%%%%%
\subsection{The Cosmic Distance Scale, Dark Matter and Dark Energy}
\label{sec:dist}

\noindent{\bf An IR View of Periodic Variables in the Local Universe}:
Pulsating variable stars (WP: \href{http://www8.nationalacademies.org/astro2010/DetailFileDisplay.aspx?id=381}{Walkowicz}) are preeminent distance indicators in the local universe.
When RR Lyrae\index{Variables!RR Lyrae} and 
Cepheids\index{Variables!Cepheids} are used, however, unmodeled dust\index{dust} and
metallicity effects manifest directly as uncertainties in
cosmological parameters \citep{bono03,tammann08}. 
SASIR would allow for precise calibration of the period--luminosity 
(P-L)\glossary{name={P-L},description={Period-Luminosity}} relation of
these two classes in the IR, largely skirting the problems with dust (and potentially metallicity \cite{scv06}) which 
complicate similar efforts at optical wavebands. 
Multiple observations at random phases would build an
unprecedented calibration of the P-L relations and enable 
GSMTs to measure precise distances well
beyond the Local Group\index{Local Group of Galaxies} and fix the rungs of the cosmic distance ladder\index{cosmic distance ladder}
out to $\sim 25$\,Mpc.  Mira variables\index{Variables!Mira}, promising new distance
indicators, would also be observable with SASIR to $\sim$4 Mpc with
what appears to be a metallicity independent P-L relation \citep{feast08}.

\begin{figure}[hbt]
	\renewcommand{\baselinestretch}{0.92}\normalsize
	\begin{minipage}{2.0in}
\caption{{\footnotesize \it Expected SN counts for two types of surveys
    with SASIR: a rolling search with weekly cadence (left), and
    ``all-sky'' monitoring every two months (right). The number of SNe
    in the different surveys is inversely proportional to the
    light-curve information we will have for each SN. The samples are
    complementary and both have cosmological use. \label{fig:SNe}}}
\end{minipage} \hspace{0.025in} \begin{minipage}{4.7in}
\centerline{\includegraphics[width=2.36in,height=2.13in,angle=0]{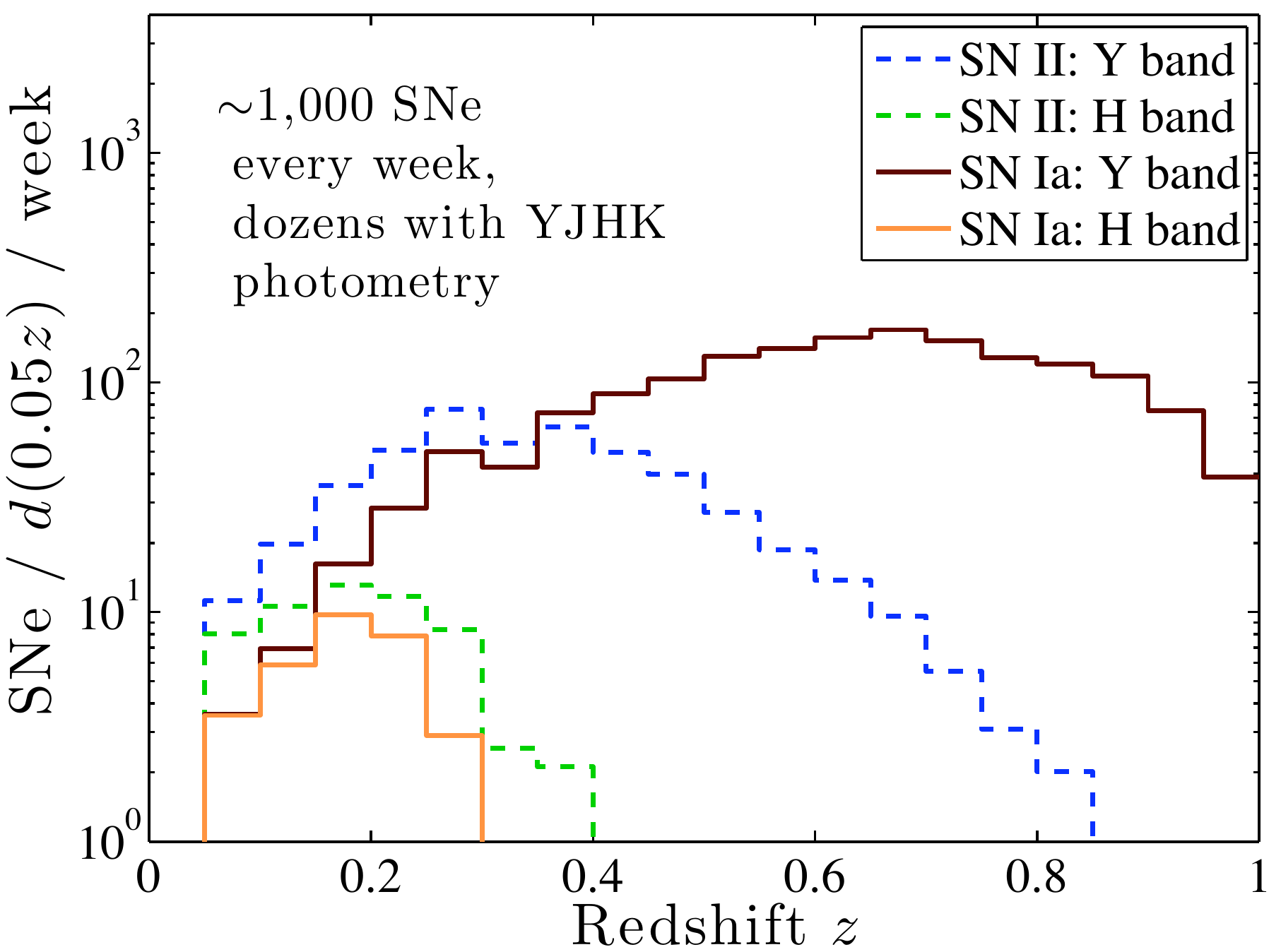}
\includegraphics[width=2.33in, angle=0]{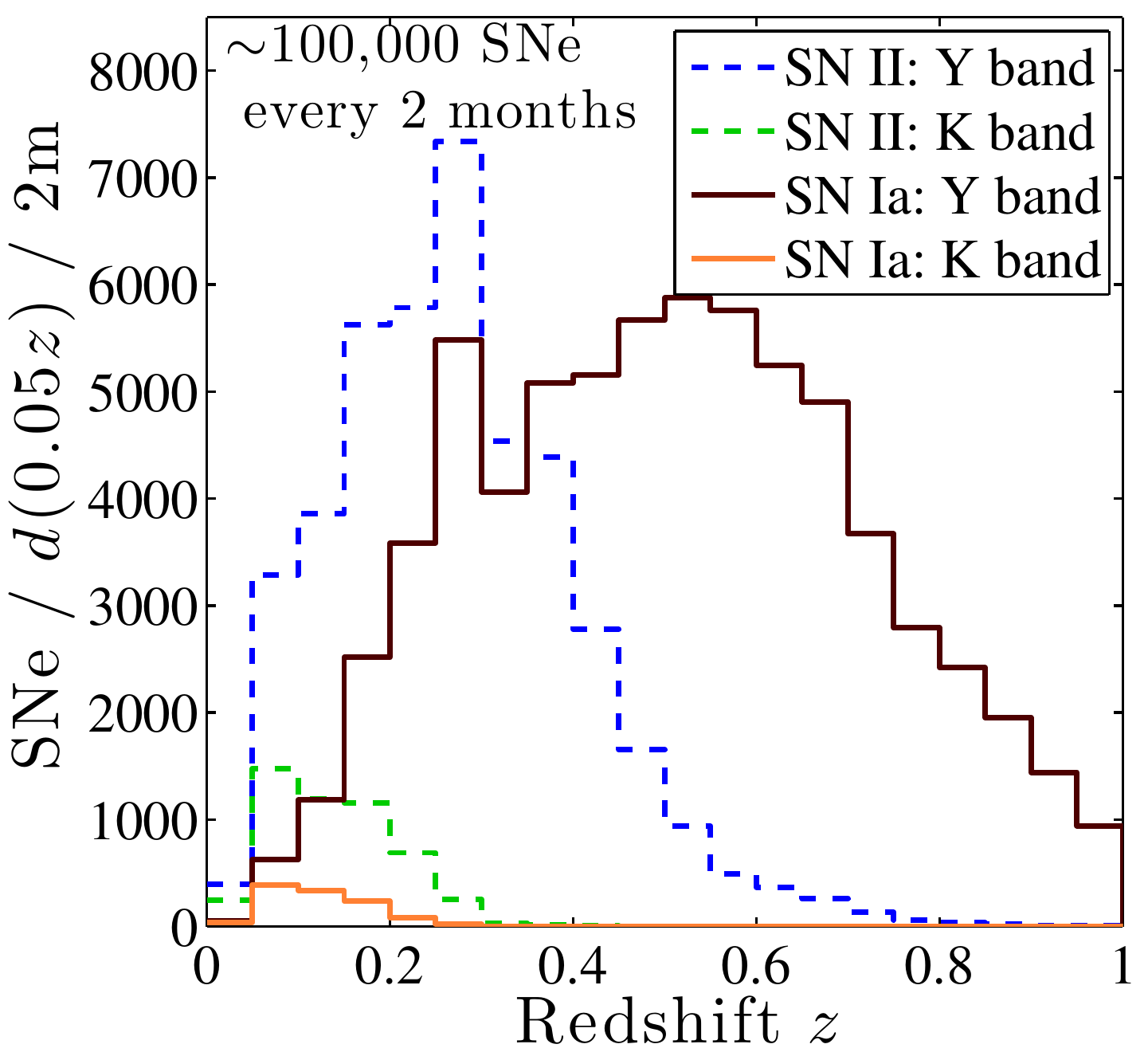}}
\end{minipage}
\end{figure}

\noindent {\bf Infrared Supernovae:} 
Type Ia supernovae\index{Supernovae}\index{Supernovae!Ia} (SNe\,Ia) (WP: \href{http://www8.nationalacademies.org/astro2010/DetailFileDisplay.aspx?id=365}{Howell}) \glossary{name={SNe},description={Supernovae}} are
better standard candles in the IR than at optical wavebands
\citep{wood-vasey07b} and minimize systematic effects that plague Ia
optical cosmography\index{Cosmography}. Type II-P\index{Supernovae!II-P} SNe are an emerging distance indicator (e.g., \cite{1992ApJ...395..366S}) recently shown to be almost comparable in precision to SNe\,Ia (\cite{poznanski08}; WP: \href{http://www8.nationalacademies.org/astro2010/DetailFileDisplay.aspx?id=150}{Poznanski}), and could be
even better standard events in the IR. 
In addition to cosmography, searching
for SNe in the IR will improve our understanding of 
their rates (largely circumventing dust\index{dust} effects) and
hence constrain their diverse progenitors. Using optically determined rates alone, SASIR will detect more than 10$^6$ SNe during the four-year survey (Fig.\ \ref{fig:SNe}).

\noindent{\bf Photometric Redshifts for Weak Lensing and Baryonic
  Oscillations}: Wide-field IR photometry provides a compelling
improvement in the photo-$z$ measurements of galaxies, especially
beyond $z \sim 1.5$ (Fig.\ \ref{fig:photo}, left). 
\index{Weak lensing}\index{Baryonic Oscillations} 
The enhanced accuracy for
galaxies across the Northern sky will greatly improve the returns on
weak lensing and baryonic oscillation experiments with wide-field
optical facilities (WP: \href{http://www8.nationalacademies.org/astro2010/DetailFileDisplay.aspx?id=387}{Riess}, \href{http://www8.nationalacademies.org/astro2010/DetailFileDisplay.aspx?id=208}{Eisenstein}, \href{http://www8.nationalacademies.org/astro2010/DetailFileDisplay.aspx?id=200}{Heap},
\href{http://www8.nationalacademies.org/astro2010/DetailFileDisplay.aspx?id=309}{Zhan })

\noindent{\bf Galaxy Evolution and High-redshift Clusters}: A
frontier endeavor for the next decade will be to determine the
progress of nascent galaxies as a function of local environment as
they proceed from the ``blue swarm'' of small star-forming objects at
$2 < z < 3$ to the well-defined red sequence of massive galaxies seen
in both galaxy clusters and the field at $z < 1$ (WP:  \href{http://www8.nationalacademies.org/astro2010/DetailFileDisplay.aspx?id=123}{Holden}, \href{http://www8.nationalacademies.org/astro2010/DetailFileDisplay.aspx?id=372}{Labbe}, \href{http://www8.nationalacademies.org/astro2010/DetailFileDisplay.aspx?id=382}{Stanford}).  The appearance of
the red sequence will probably spread from the high- to low-density
environments, such as groups, before becoming established in the
field.  Such a study requires identifying the full range of
environments at $z > 2$, through wide-area imaging, since halos of mass
$M > 10^{14}$\, M$_\odot$ are exceedingly rare at $z > 2$ (Fig.\
\ref{fig:photo}, right).

\begin{figure}[tbh]
	\renewcommand{\baselinestretch}{0.92}
\centerline{\includegraphics[width=2.6in,angle=0]{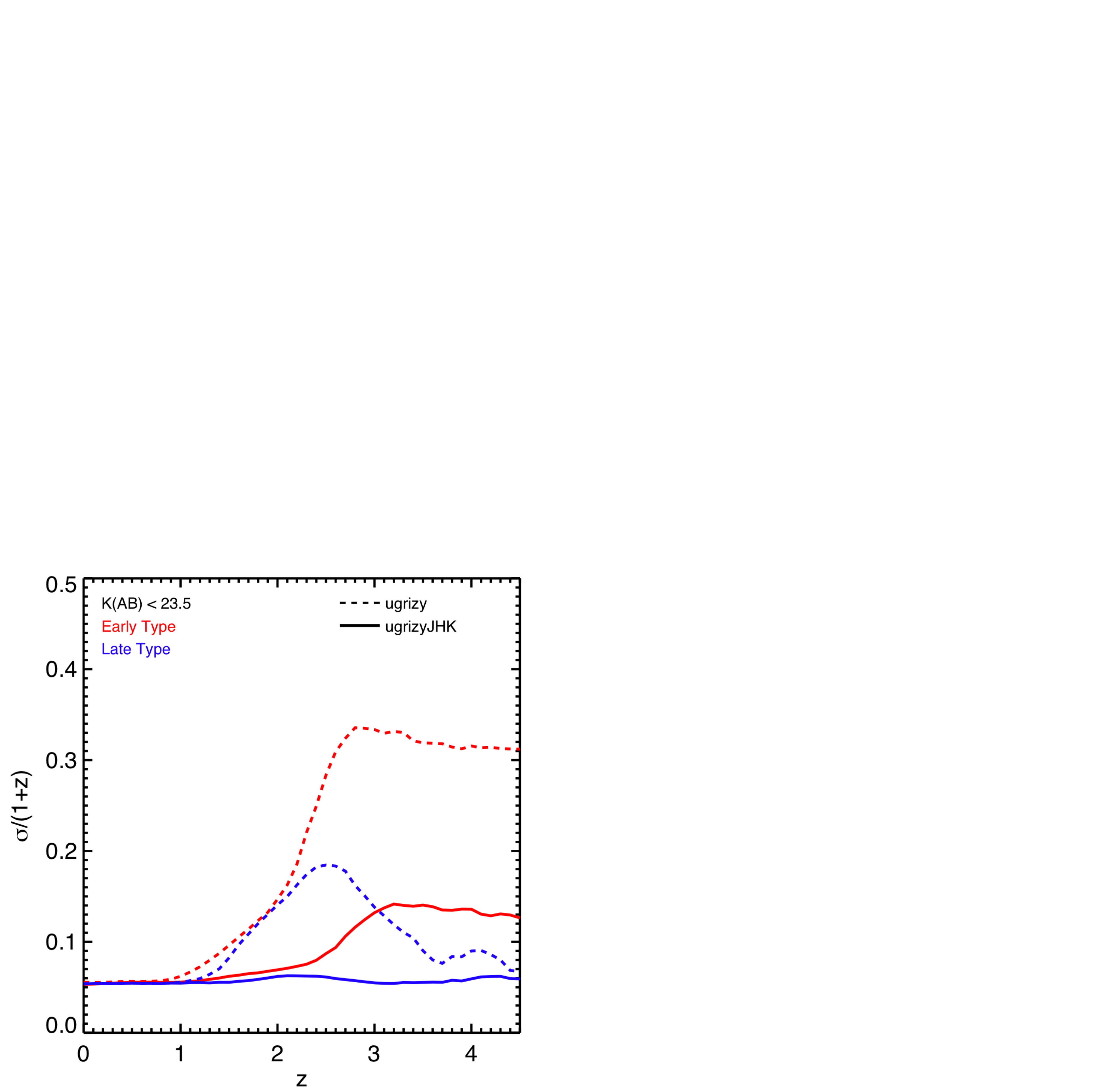}
\includegraphics[width=2.6in, angle=0]{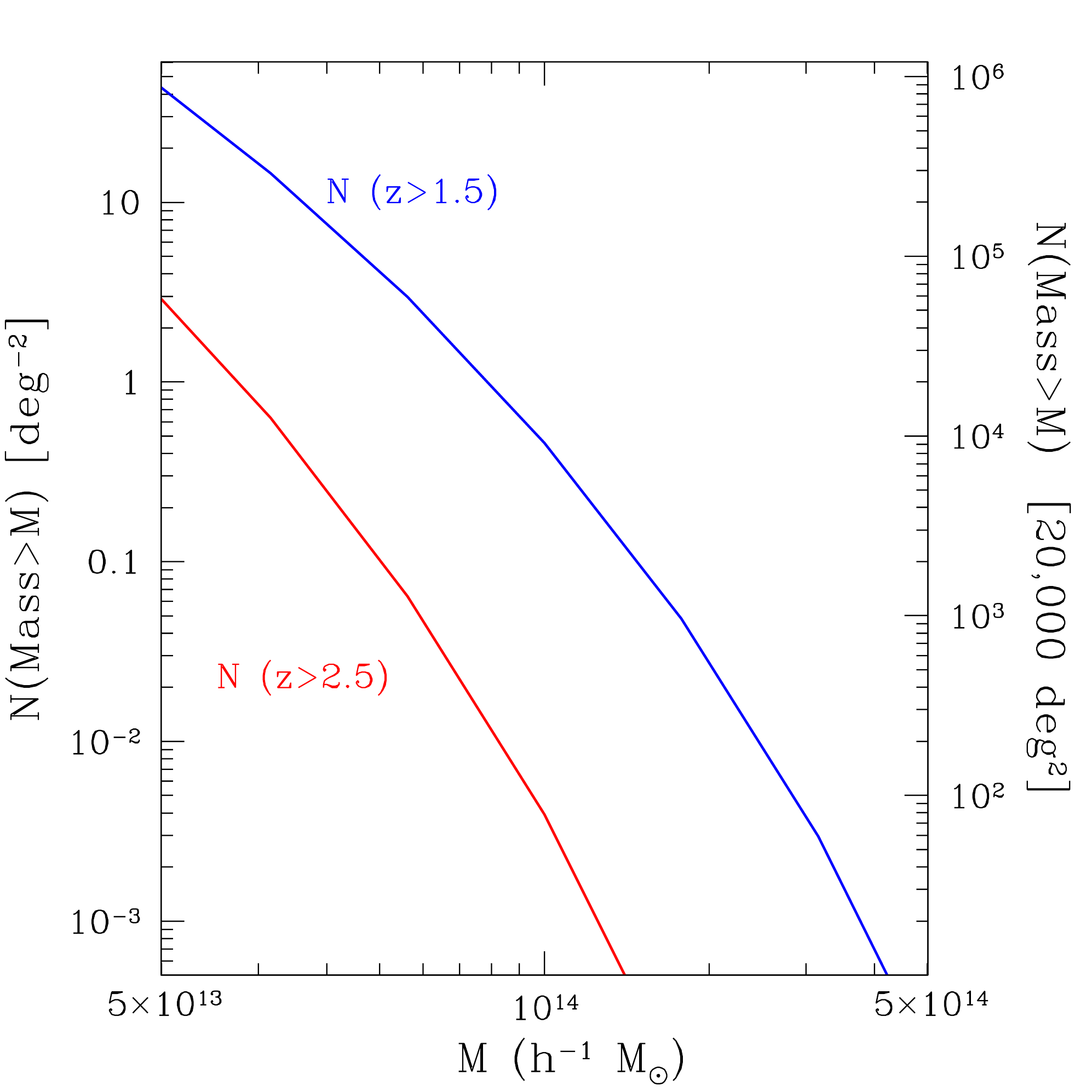}}
\caption{{\footnotesize \it (Left) Simulated photometric redshift \index{Photometric redshift}
    accuracy for early (red) and late-type (blue) galaxies. Dashed
    lines show the expected performance of LSST/Pan-STARRS4\index{Pan-STARRS}\index{LSST}
    alone. Solid lines include NIR data from SASIR, which vastly
    improve accuracies at $z >1$. Simulations assume a 4\% floor to
    these accuracies \cite{2006ApJS..162...20B}. (Right) Expected cluster
    demographics by mass in two redshift regimes. Finding the rarest
    massive clusters requires a wide-field near-IR imaging survey so
    that galaxy populations at $z > 2$ are selected at rest frame
    wavelengths. \label{fig:photo}}} \vspace{-0.2in}

\end{figure}

\vspace{-0.1cm}
\subsection{A New Phase Space for Transient Discovery}
\label{sec:gw}

\vspace{-0.05cm}

\index{Transients}

\begin{figure}[t]
	\renewcommand{\baselinestretch}{0.92}
	\begin{minipage}{3.1in}
	\caption{  \it \footnotesize  	
Characterizing the transient\index{Transients} universe at IR wavelengths. Aside from the ``known'' Type Ia (blue) and core-collapse (red) SNe, new types of extragalactic transients are expected to arise from cataclysmic events. Shown are our estimated restframe infrared light curves resulting from the collision between a NS\index{Neutron star} and red supergiant (RSG; purple), the disruption and ignition of a white dwarf by an intermediate mass BH (yellow), and the merging of a NS binary (powered by r-process nucleosynthesis; black). SASIR will readily see NS-NS\index{Neutron star mergers} mergers to the Advanced LIGO volume (see $\S$~\ref{sec:gw}). These preliminary model calculations suggest that an assortment of peculiar transients should be uncovered by SASIR, providing a complementary view (less obscured by dust\index{dust}) of the transient universe than that offered by optical synoptic surveys. 	\label{fig:transients}}
\end{minipage} \begin{minipage}{3.3in}
\centerline{\includegraphics[width=3.3in]{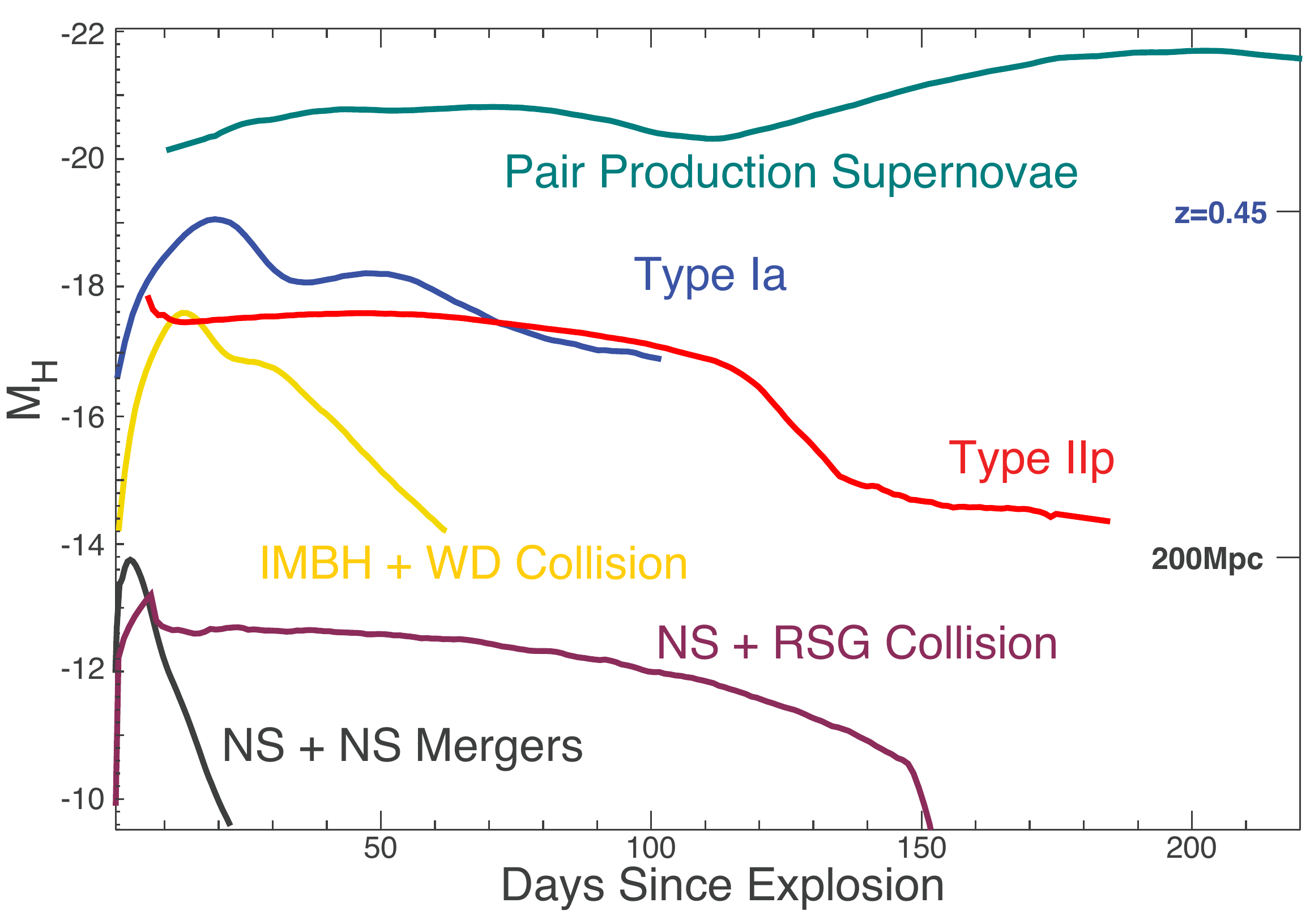}}
\end{minipage}
\end{figure}

\glossary{name={EM},description={Electromagnetic}}  
Deep multicolor synoptic monitoring on hundreds to thousands of square degrees on minutes to months timescales would break new ground in the infrared, opening up the potential for
totally new classes of objects found by IR variability. There are indications that exploration in this space phase will be fruitful (e.g., WP: \href{http://www8.nationalacademies.org/astro2010/DetailFileDisplay.aspx?id=191}{Kulkarni} \href{http://www8.nationalacademies.org/astro2010/DetailFileDisplay.aspx?id=281}{Wozniak}, \href{http://www8.nationalacademies.org/astro2010/DetailFileDisplay.aspx?id=370}{York}), particularly relevant to ``multi-messenger'' astrophysics\index{Multi-messenger astrophysics}. Indeed the explosive events which dominate the high-energy sky --- \index{Gravity waves}  involving compact objects such as neutron stars\index{Neutron Stars} (NSs) and BHs\index{Black Holes} (BHs) (of
both the stellar and supermassive varieties) --- should produce long-wavelength signatures (Fig.\ \ref{fig:transients}). Motivating the need for wide-field monitoring, TeV gamma-ray \v{C}erenkov \index{TeV Gamma-ray Telescopes} telescopes and
neutrino detectors will localize events only to degree-scale
accuracy. Likewise, Advanced LIGO\index{LIGO} and LISA\index{LISA} are
expected to localize degenerate object merger events through
gravitational waves 
(GW)\glossary{name={GW},description={Gravitational wave}} with, typically, 
one-degree scale
uncertainties.  Much of the science extracted from these new windows
on the universe will require the identification of electromagnetic
counterparts, which would yield the redshift of the host galaxy and
enable their use as standard sirens for cosmography
\citep{schutz86,kocsis08}\index{Cosmography} (WP: \href{http://www8.nationalacademies.org/astro2010/DetailFileDisplay.aspx?id=18}{Bloom}). Little is
known of these signatures but indications are that the IR is a valuable window \citep[Fig.\ 5;][]{2008ApJ...684..835S,2008arXiv0807.4697H} (WP: \href{http://www8.nationalacademies.org/astro2010/DetailFileDisplay.aspx?id=58}{Hawley}, \href{http://www8.nationalacademies.org/astro2010/DetailFileDisplay.aspx?id=350}{Phinney}, \href{http://www8.nationalacademies.org/astro2010/DetailFileDisplay.aspx?id=18}{Bloom}).  
As such, SASIR, with its rapid access to deep wide-field imaging is a
promising tool.

\vspace{-0.13cm}
\subsection{Broad Scientific Reach and Synergies with Other Major Facilities}

\vspace{-0.06cm}

The above discussion highlights the expected impact of SASIR in just some of the
fields of interest in the next decade. Many of these areas were
covered in the \href{http://sasir.org/whitepapers2}{whitepapers
  generated by our collaboration}. The following lists additional
examples with hypertext links to science WPs submitted by us and
other groups to Astro2010:

\vspace{0.08in}

- Discover new satellites of the Galaxy minimizing dust biases (WP: \href{http://www8.nationalacademies.org/astro2010/DetailFileDisplay.aspx?id=296}{Bullock}, \href{http://www8.nationalacademies.org/astro2010/DetailFileDisplay.aspx?id=308}{Johnston}),

- Produce large new samples of strong lenses (WP: \href{http://www8.nationalacademies.org/astro2010/DetailFileDisplay.aspx?id=375}{Coe}, \href{http://www8.nationalacademies.org/astro2010/DetailFileDisplay.aspx?id=205}{Koopmans}, \href{http://www8.nationalacademies.org/astro2010/DetailFileDisplay.aspx?id=282}{Marshall}),

- Survey the oldest (i.e.\ coldest) white dwarfs in the
Milky Way (WP: \href{http://www8.nationalacademies.org/astro2010/DetailFileDisplay.aspx?id=292}{Kalirai}),

- Produce a Galactic dust-extinction map of unparalleled spatial resolution
(WP: \href{http://www8.nationalacademies.org/astro2010/DetailFileDisplay.aspx?id=221}{Gordon}),

- Probe the bright end of the galactic luminosity function
at $z>5$ (WP: \href{http://www8.nationalacademies.org/astro2010/DetailFileDisplay.aspx?id=346}{Bouwens}),

- Stellar population analysis of nearby galaxies (WP: \href{http://www8.nationalacademies.org/astro2010/DetailFileDisplay.aspx?id=369}{Kalirai}, \href{http://www8.nationalacademies.org/astro2010/DetailFileDisplay.aspx?id=268}{Kirby}, \href{http://www8.nationalacademies.org/astro2010/DetailFileDisplay.aspx?id=196}{Lu}, \href{http://www8.nationalacademies.org/astro2010/DetailFileDisplay.aspx?id=100}{Meixner}, \href{http://www8.nationalacademies.org/astro2010/DetailFileDisplay.aspx?id=152}{Worthey}, \href{http://www8.nationalacademies.org/astro2010/DetailFileDisplay.aspx?id=152}{Wyse}),

- Discover star cluster systems (WP: \href{http://www8.nationalacademies.org/astro2010/DetailFileDisplay.aspx?id=230}{Rhode}),

- Study stellar morphology (bulge/disk) in nearby galaxies (WP: \href{http://www8.nationalacademies.org/astro2010/DetailFileDisplay.aspx?id=320}{Clarkson}),

- Map the distribution of low-mass stars well beyond the solar neighborhood
(WP: \href{http://www8.nationalacademies.org/astro2010/DetailFileDisplay.aspx?id=359}{Cruz}),

- The search for light bosons (WP: \href{http://www8.nationalacademies.org/astro2010/DetailFileDisplay.aspx?id=305}{Chelouche}),

- Finding Type IIn SNe at $z>5$ (WP: \href{http://www8.nationalacademies.org/astro2010/DetailFileDisplay.aspx?id=130}{Cooke}),

- Probing quasar variability (WP: \href{http://www8.nationalacademies.org/astro2010/DetailFileDisplay.aspx?id=385}{Elvis}, \href{http://www8.nationalacademies.org/astro2010/DetailFileDisplay.aspx?id=115}{Murray}),

- Long-wavelength signatures of tidal disruption events (WP: \href{http://www8.nationalacademies.org/astro2010/DetailFileDisplay.aspx?id=80}{Gezari}),

- Exploring the nature of X-ray and explosive 
    transients (WP: \href{http://www8.nationalacademies.org/astro2010/DetailFileDisplay.aspx?id=220}{Soderberg}, \href{http://www8.nationalacademies.org/astro2010/DetailFileDisplay.aspx?id=281}{Wozniak}),

- Studies of variable stars from near to far (WP: \href{http://www8.nationalacademies.org/astro2010/DetailFileDisplay.aspx?id=381}{Walkowicz})

\vspace{0.1in}
\noindent It is worth emphasizing that many of the science WPs
submitted to Astro2010 have called for wide-field, near-IR surveys of
the sky.  While some science goals demand the high spatial-resolution afforded by
space missions, most could be done for far less expense by a
ground-based observatory like SASIR.

\newpage

\section{Technical Overview}
The enormous scientific promise of SASIR is based on a dedicated
wide-field (1 degree diameter) large aperture telescope (6.5 m in
diameter), located in a dark site with a large fraction of clear
nights ($\sim 75\%$), enabling deep and synoptic imaging of the whole
Northern sky simultaneously in NIR bands ($Y$, $J$, $H$ and $K$) with four independent focal planes.

This uniquely powerful survey and facility does not require 
new technology exploration nor a totally new kind of telescope. Indeed, as described in \S \ref{sec:telescope}, the SASIR
telescope design will use already
proven and operational concepts (such as from the Magellan and MMT
telescopes) as the point of departure. Nevertheless, given the confluence of a moderately wide
field with multiple and large focal planes, its actual design and
technical feasibility with present day technology and IR materials
needs to be carefully studied in the {\it preliminary design phase} (see \S \ref{sec:sked}, the schedule of proposed activities).

\subsection{\bf The 6.5-meter SASIR Telescope}
\label{sec:telescope}

\vspace{-0.1cm}

 The telescope structure and primary mirror are to be based on the
highly successful and efficient Magellan Telescopes in operation at
Las Campa\~nas Observatory in Chile and the MMT\index{MMT} telescope at Mt.\
Hopkins\index{Mt.\ Hopkins}.  The Magellan and MMT optics and structures are each well suited to
wide-field imaging. In particular, a successful one-degree f/5
corrector has been in operation at the MMT 
\citep{Fata04}
while a second system is being commissioned at Las Campa\~nas. Furthermore, we have demonstrated  \cite{Gonz07} that this design is quite capable of delivering fields of view beyond 1.5$^\circ$ in diameter. 
The three main challenges facing the design of the SASIR telescope are therefore of a different nature:

\begin{itemize}

 \item Simultaneous feeding of up to four focal planes, with collimator/camera 
  NIR optics of reasonable size;

 \item Controlling spurious thermal emission in the $H$ and $K$ bands (under a proper baffling 
     system, coupled with space for a cold pupil);

 \item Maintaining the instrument within the weight and envelope limits of the Magellan or MMT
       structure.

\end{itemize}

\begin{figure}[tb] 
\centerline{ 
\includegraphics[width=6.5in]{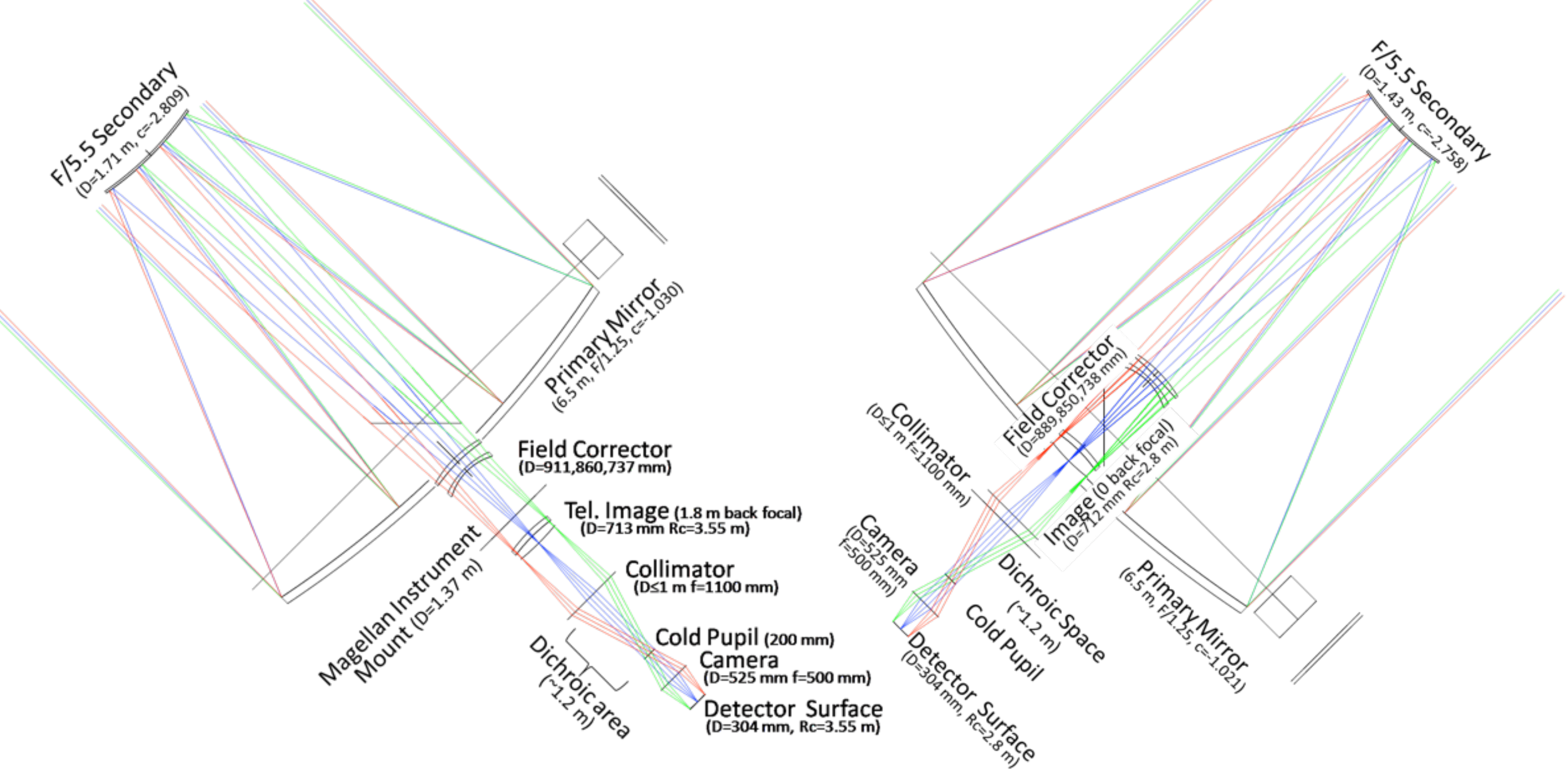}}
\caption{\it SASIR telescope designs (Magellan nominal back
focus at right, null back focal at left). A single arm is shown with
indicative camera and collimator\index{collimator} parameters (folding dichroics not
drawn). The telescope and corrector deliver and effective f/5.5 focal,
the whole system is $\sim$f/2.5. Relevant characteristics of the main
components are shown. These two baselines will be studied (optics,
mechanics, costing) in detail.}
\label{fig:OptDes}
\end{figure}

\index{SASIR!Telescope}
The first challenge above not only drives the survey
speed but makes SASIR different and quite powerful with respect to its
closest relatives, the 4m-class VISTA\index{VISTA} and UKIDSS\index{UKIDSS} systems and the 8m
LSST. During 2009, the SASIR collaboration will be developing its
telescope concept by fully investigating and resolving among potential
telescope solutions that optimize the science returns and minimize the
risks of the camera design: a conventional telescope plus image
reducer(s), as here presented, a 3-mirror telescope \index{3-mirror telescope design} with an
intermediate collimated beam, or a conventional telescope with
dichroic(s) \index{dichroic} in non-parallel beams.

A range of variants on the base Magellan design are currently under
consideration to find a solution that will permit up to four
individual-band cameras, each with $\sim$1$^\circ$ field of view (FoV), with NIR refractive
optics under about 500 mm in diameter.  The current reference concept
consists of a f/5.5 Ritchey-Chretien design with a 3-lens field
corrector (all spherical, Silica-like glass).  In order to allow for a
cold pupil as well as the placement of dichroics, the \index{dichroic} telescope is
coupled to a focal reducer, with a collimated beam of 200 mm and a
camera with focal length of 500 mm. Figure~\ref{fig:OptDes} shows two
examples of the telescope concept, the first one maintaining the back
focal distance of Magellan, while the second telescope focuses at the
primary vertex, exploring the range in which the diameter of the
secondary and the height of the pupil can be controlled. The present
studies include cases for telescope f-ratios from f/3.5 up to f/11, at
both Cassegrain and Nasmyth stations, and a range of pupil
diameters. The designs shown deliver an image quality close to the
diffraction limit across the whole FoV, between 0.03$''$ and 0.1$''$
FWHM.  These idealized (pre-construction) telescope performances
indicate that most of the optical error budget can be left for the
more difficult collimator and camera designs, as well as for the
construction and operation of the entire system. These telescope
concepts let us explore the main parameter space and general dimensions for the SASIR collimator and camera systems.  The full range of parameter exploration expected during the conceptual and preliminary design phases is detailed in Table \ref{tab:study}.

%\vspace{-0.14in}

\vspace{-0.15cm}
\subsection{The SASIR Camera}

\noindent {\bf Foreoptics}: SASIR plans a split-beam design for the camera\index{SASIR!Camera} optics, 
like 2MASS, to simultaneously image in 4 filters. The full optical
design of SASIR will be driven by the following guidelines and
constraints: the aperture and curvature of the primary mirror
(f/1.25), a FOV of 1.06$^circ$, 
a plate scale of
$\approx$~0.228$''$ per 18~$\mu$m pixel (f/2.5 net system), an
after-construction-under-operation image quality that does not
deteriorate by more than a few percent the median NIR seeing, a high-throughput design (e.g.\ efficiency $>30\%$) at least within the $\lambda = $ 0.8--2.4~$\mu$m
range, and a system with low thermal emission from its optical
components that also permits the proper buffering of scattered thermal
emission. The conceptual designs of the collimator and camera, based
essentially on a scaled version of already known systems (e.g.\
FourStar\index{FourStar} NIR Camera for the Magellan Telescope; \cite{2008SPIE.7014E..95P}), will be
developed in parallel and share optimization constraints with the telescope
concept.

\begin{figure}[tb] 
\centerline{ 
\includegraphics[height=5.3in,angle=270]{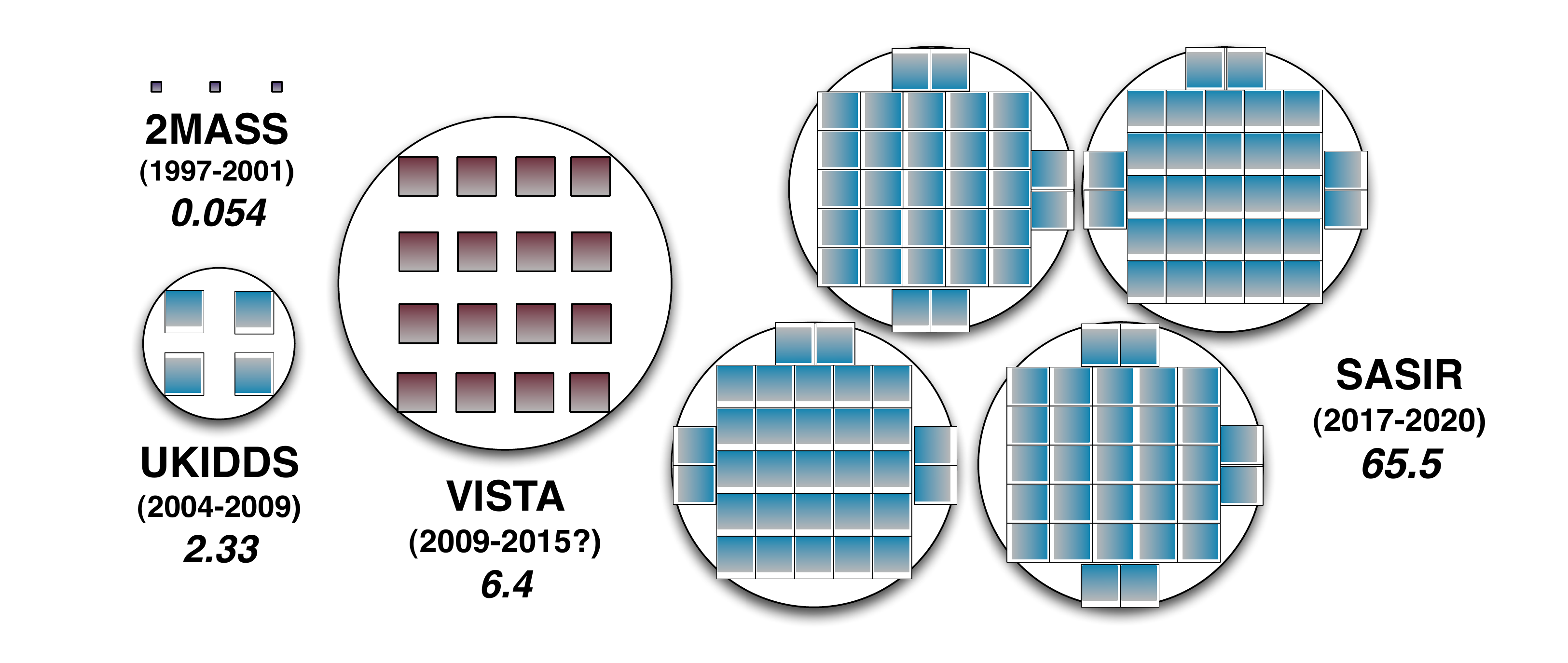}}
\caption{\it To scale physical comparison of the focal planes of 2MASS, UKIDDS, VISTA and SASIR. The {\'entendue-couleur} (m$^2$ deg$^2$ $\times$ number of simultaneous bands), the instantaneous light grasp, is shown for each facility.\index{2MASS}\index{VISTA}\index{UKIDSS}}
\label{fig:focal}
\end{figure}

\noindent {\bf Detectors}: Given the expense for science-grade IR detectors (\S \ref{sec:cost}), 
our design is driven by a desire to cover a large field of view with
the fewest pixels while still adequately sampling the good seeing at
SPM.  The nominal detectors are 2048 $\times$ 2048
arrays with 18 or 20 $\mu$m pixels, now commercially
available.\footnote{The SASIR collaboration has built a strong working
relationship with the two main vendors of IR arrays (Teledyne\index{Teledyne} and
Raytheon) \index{Raytheon}  and we have gained confidence that the necessary volume of
arrays can be delivered on a timescale shorter than the telescope
construction time.} 
This translates to 0.228 -- 0.25 $''$/pixel. We
are baselining 124 science-grade arrays
(Figure \ref{fig:focal}). SASIR will not be sensitive to the
read-noise of the detectors, as even short exposures are expected to
be background limited. We will thus allow for as flexible and dynamic
imaging as the science requires.

\vspace{-0.15cm}

\subsection{\bf Survey Strategy and Cadence Optimization}
\vspace{-0.1cm}

The Survey optimization will be revised in detail during the {\it
preliminary design}, accounting for the articulated priorities of the
diverse science cases. The baseline plan calls for 20 second
double-correlated exposures -- this optimizes on-sky exposures without
saturating fainter 2MASS\index{2MASS} stars (which will be crucial for establishing
the photometric baseline). For a total on-source dwell of 80 seconds
(2 visits per night consisting of 2 integrations each) and nominal
slew time to next field of 6 seconds, we expect to cover about 140
sq.\ deg per 8-hour night, implying that the entire visible sky from a
single site could be imaged every 2--3 months. Table~1
shows the expected point source and extended source sensitivities (see also Figure~\ref{fig:comparison}).
The simplest survey strategy would be to cover the sky
repeatedly with roughly equal time between visits. Over a 4 year
survey, each position could be observed $\sim$6 times. To determine
the parallax and proper motion of objects in the solar neighborhood
(\S \ref{sec:lm}), we require at least three visits per field.  In
practice, there will be several different cadence strategies, with
both competing and complementary goals. For instance, a fast transients search would yield very deep imaging in several hundreds of degrees squared. The SN\index{Supernovae}
search (\S \ref{sec:dist}) would benefit from repeated scans of the
same part of the sky every few nights, while a search for high proper
motion objects would only require repeat observations on a
months to years timescale.

\begin{deluxetable}{lcccccc} \label{tab:sens}
\tablewidth{0pt}
\tabletypesize{\small}
\tablecaption{Nominal Sensitivities from SASIR Concept Design}
\tablehead{\colhead{} & \multicolumn{4}{c}{\textbf{Point Source Sensitivity}}& \multicolumn{2}{c}{\textbf{Extended Source Sensitivity}} \\
\multicolumn{1}{c}{Filter} & \multicolumn{2}{c}{Single Epoch (5-$\sigma$)} & \multicolumn{2}{c}{Survey (5-$\sigma$)} & \multicolumn{2}{c}{Survey (5-$\sigma$ per pixel)} \\ 
\colhead{} & \colhead{[AB mag]} & \colhead{[$\mu$Jy]} & \colhead{[AB mag]} & \colhead{[$\mu$Jy]} & \colhead{[AB arcsec$^{-2}$]} & \colhead{[$\mu$Jy arcsec$^{-2}$]}}

\startdata
Y            & 23.49 & 1.45 & 24.47 & 0.59 & 23.32 & 1.71 \\ 
J            & 22.95 & 2.40 & 23.93 & 0.97 & 22.78 & 2.82 \\
H            & 22.60 & 3.30 & 23.57 & 1.35 & 22.42 & 3.89 \\
K$_{\rm s}$  & 22.47 & 3.74 & 23.44 & 1.52 & 22.29 & 4.40 \\
\enddata
\label{photometry}
\vspace{-0.3in}\begin{spacing}{0.90}
\tablecomments{\footnotesize 
 Based on a preliminary simulation of a four band survey ($Y$ and 2MASS filters $J, H, Ks$) with 75\% clear weather fraction and average seeing of 0.6 arcsec and 18 $\mu$m pixels. Each epoch assumes 80 sec total integration with 6 epochs per field over the entire survey (24,000 sq.\ deg.). As a consistency check to the simulation, note that the $5 \sigma$ limiting magnitude of 2MASS\index{2MASS} (1.3m diameter, 7.8 s integration, seeing $\sim$~2.5\arcsec) was 17.55 AB (346 $\mu$Jy). For sky-limiting imaging, the limiting magnitude increases as 2.5 $\log_{\rm 10}$ (diameter time$^{0.5}$/seeing), with diameter, time, and seeing expressed as ratios. For the nominal six visit all-sky survey, this implies a nominal depth of 5.4 mag fainter than 2MASS. }\end{spacing}
\end{deluxetable}

\begin{figure}[tbh]  
        \centerline{\includegraphics[width=6.3in]{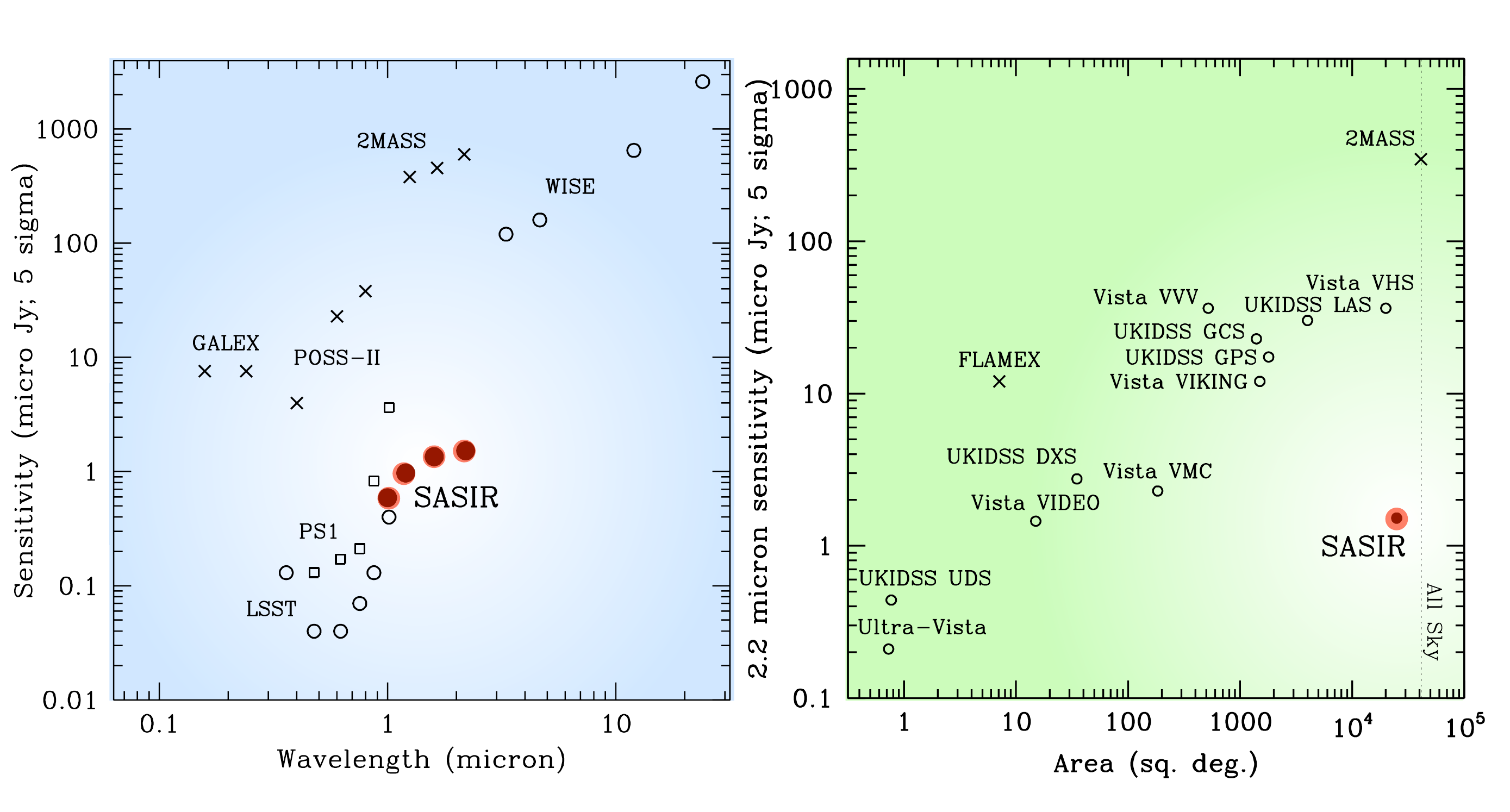}}
\caption{\small {\it Comparison of the nominal SASIR survey with other significant surveys already completed ($\times$  symbol) or planned ($\circ$ or square symbol). Left: The point source \index{SASIR!Point source sensitivity} sensitivity versus wavelength for wide-area surveys where we have assumed 6 total visits (480\,s) for SASIR. Right: The point source $K$-band sensitivity versus sky coverage.  The other survey data for these figures were compiled by D.\ Stern (JPL).}} 
\label{fig:comparison} 
\end{figure} 

\vspace{-0.1cm}

\subsection {\bf Data Taking and Data Management Strategies}

\vspace{-0.1cm}
SASIR Data Management will benefit from direct experience with the Peters Automated
Infrared Imaging Telescope (PAIRITEL) \index{PAIRITEL} Project \citep{bsb+06}, the
largest time-domain robotic telescope operating at infrared
wavelengths. In particular its low-level telescope and camera
interfaces, pipeline and archive system, and autonomous scheduling
system are a reference for determining a baseline datataking strategy
commensurate with instrument limitation and SASIR science goals. We envision that
the Data Management architecture, developed fully in the conceptual design phase (\S \ref{sec:sked}), will most resemble that of Pan-STARRS\index{Pan-STARRS}, UKIDSS and VISTA.

\vspace{-0.1cm}

\subsection{Project Site and Required Infrastructure}

\vspace{-0.1cm}

The site, located in the northern part of Mexico in the Sierra San
Pedro M\'artir in the state of Baja California\index{Baja California} at an altitude of
2890 m, has been developed over the last forty years and has three
telescopes with main optics diameters of 2.1-, 1.5-, and 0.84-m, with
a number of photometric, spectroscopic and imaging capabilities in the
optical, near-, and mid-infrared regimes. SPM\index{SPM} excels in the
transparency and darkness of the night sky as well as in the seeing
quality and stability \cite{cg03,tca07,tap07}. Comparison with
other sites suggests 
that SPM has the largest percentage of clear
nights of any site in the Northern Hemisphere. The median seeing
reported by \citealt{michel03} is $0.6''$ in the $V$ band. It is yet to host
a competitive next-generation telescope facility that fully exploits
its unique virtues. The site had been considered by the LSST
consortium prior to the decision to locate it in Cerro Pach\'{o}n
(Chile)\index{Cerro Pach\'on}, as well as by next-generation extremely large telescope
projects, such as the \index{Giant Segmented Mirror Telescopes!Thirty Meter Telescope} Thirty Meter Telescope (TMT). \glossary{name={TMT},description={Thirty Meter Telescope}}

 The road that climbs the Sierra is paved up to the entrance to the
National Park, while the last 16 km (within the park) 
is being completed now.  Ample lodge and
workshops at the Observatory are located 2 km from the telescopes and
the selected SASIR\index{SASIR!Site} site. The Observatorio Astron\'omico Nacional \glossary{name={OAN},description={Observatorio Astron\'omico Nacional}} (OAN)\index{OAN} has five electric generators with
capacities of 280, 230, 200, 150, and 90 kw, respectively, but plans
to soon connect SPM to the public electric network. 
This new line will
also carry a fiber-optic network for a very high speed network,
upgrading the current microwave link. 
IA-UNAM has a
research branch in nearby Ensenada\index{Ensenada} with a staff of 25
astronomers and a similar number of technical staff and
students. IA-UNAM-Ensenada \index{IA-UNAM} also serves as the logistic and
administrative station for the observatory at SPM. Given the state of
the present and planned infrastructure on the site, SASIR does not
require, other than the facility itself, further major construction at
SPM.

\vspace{-0.1cm}

\subsection {Design Considerations for the 2nd-Phase Operations}
The SASIR Telescope is expected to be operational for 30+ years, well
after the SASIR Survey is completed. The consortium envisions the
telescope to continue mostly as a dedicated surveying facility. The current plan is to perform a wide-field optical/NIR spectroscopic
survey, making use of the large detector investment. The SASIR telescope will be designed so as not to 
preclude or undermine the feasibility of such a multiobject spectroscopic survey.

\newpage
\section{Technology Drivers}

\bigskip

\bigskip

While the construction of a 6.5m telescope and a wide-field 4-channel IR camera will be an ambitious undertaking, we have not identified any truly ``new technologies'' that would need to be proven in advance of construction. To be sure, there are significant technical challenges and risks will need to be retired appropriately (see \S \ref{sec:cdc}). For example, it is unclear that current optics vendors
can produce sufficiently large, IR transmissive optics for the proposed wide-field cameras.  Such concerns will be addressed in the  design phases and may be mitigated by other (larger-scale) projects that have similar needs, e.g. GSMT\index{Giant Segmented Mirror Telescopes}.

\newpage
\section{Activity Organization, Partnerships, and Current Status}
\vspace{-0.15in}
\noindent{\bf Project History \& Current Status}: The SASIR initiative started in late 2007 and the design is still at a {\it preliminary conceptual level}. No significant funding for the project itself has been granted, though several major proposals for support of the design phase were solicited in both the US and Mexico starting in Fall 2008. The preliminary conceptual design was created in tandem with the development of the SASIR science case over 2008, culminating in the production of a preliminary whitepaper by the collaboration for dissemination to members of the institutional partners. Two principals meetings were conducted in the first half of 2008 (in Santa Cruz and in Mexico City). A \href{http://www.inaoep.mx/~progharo/gh2008/}{two-week collaboration workshop}, with $>$40 people in attendance, was held in Puebla, Mexico in August 2008. The most significant engagement (face-to-face meetings, regular telecons) with third party vendors to-date has been with detector manufacturers (Raytheon and Teledyne). The casting of the primary mirror is currently underway.

\noindent{\bf Institutional Partners}: SASIR is an international partnership between the University of California, Instituto de Astronom\'ia at Universidad Nacional Aut\'onoma de M\'exico (IA-UNAM) \glossary{name={IA-UNAM},description={Instituto de Astronom\'ia at Universidad Nacional Aut\'onoma de M\'exico}}\index{IA-UNAM}\index{UNAM}, the Instituto Nacional de Astrof\'isica, \'Optica y \'Electronica (INAOE)\glossary{name={INAOE},description={Instituto Nacional de Astrof\'isica, \'Optica y \'Electronica}}, and the University of Arizona. The \index{SPM}\index{INAOE} SPM site (OAN) is operated by UNAM. The University of Arizona (developer of the Magellan/MMT mirrors)  was commissioned  by INAOE for the casting and figuring of the primary mirror. INAOE and the University of Arizona co-share ownership of the primary mirror. Extending partnerships, particularly to US and Mexico national funding agencies \glossary{name={NSF},description={National Science Foundation}} (CONACyT, NSF, etc.), is a priority for \glossary{name={CONACyT},description={Consejo Nacional de Ciencia y Tecnolog\'ia}} the collaboration. Other university or national laboratory partners may be solicited pending the outcome of private fundraising activities (see \S \ref{sec:cost}).

\noindent{\bf Current Project Management}: All work on SASIR has been through in-kind contributions by the various principals involved in the project. The diagram below shows the current structure of the collaboration as we finish the {\it preliminary conceptual design} phase. 
\begin{figure}[h] 
\centerline{ 
\includegraphics[width=5.5in,height=1.5in]{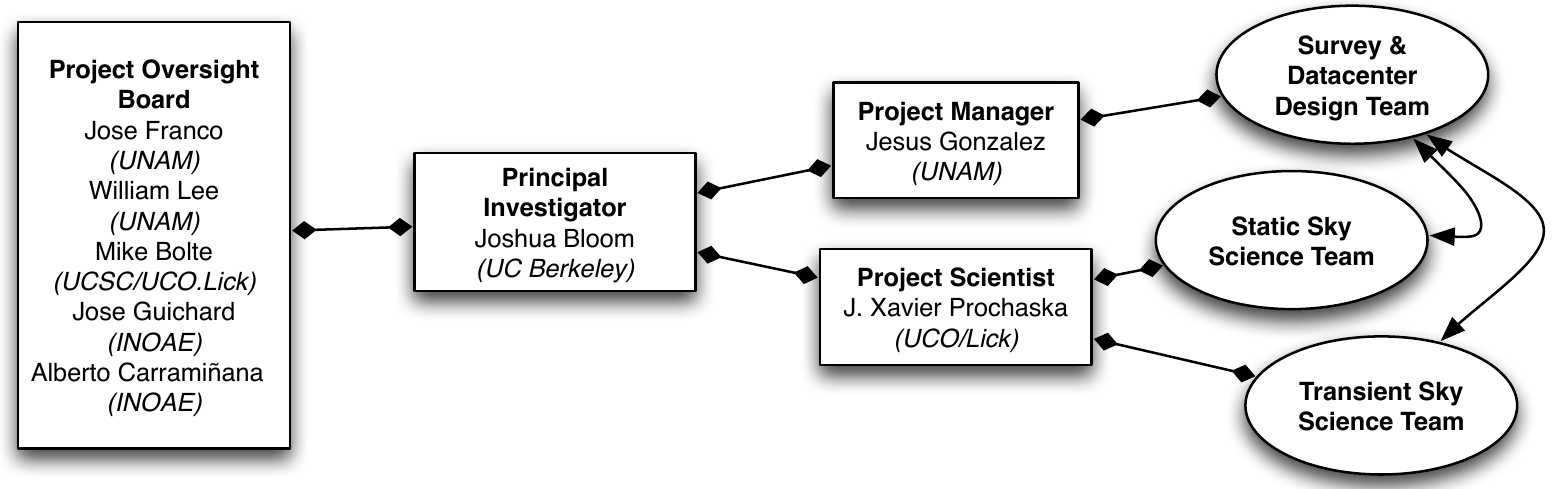}}
\end{figure}
A reorganization (and expansion) of the management structure is expected with the first round of funding for the {\it conceptual design}.

\noindent{\bf A Pathfinder Instrumentation Project:}
Our collaboration has begun construction of a new camera, the Reionization \& Transients Infrared (RATIR)\index{RATIR} Camera, to be housed on the 1.5m telescope at SPM. The 2-year experiment has its own transients science drivers, but will serve as an important pathfinder for SASIR development: RATIR will be used for collaboration building, to engage 3rd party vendors and to help the collaboration gain first hand experience with the detector operations in advance of SASIR. RATIR will obtain nightly transmission and sky brightness statistics in $Y$, $J$, $H$ for the duration of that experiment; this will directly feed into the SASIR survey simulations. As of now, we have only limited information about the SPM\index{SPM} IR sky background as a function of lunar phase and almost no information about the $Y$-band site metrics.

\newpage
\section{Activity Schedule}
\label{sec:sked}
\vspace{-0.2in}
The SASIR telescope and camera are to be developed over an eight year
span, starting in 2009. The facility will be operated to carry out and
complete the SASIR Survey in four to five years,
starting in 2017. The schedule for the SASIR activities can be
summarized in the following schematic phases (Figure~\ref{fig:gantt}), 
with particular aims and terminal points:

\begin{enumerate}
 \item {\bf Project Establishment (2009)}: a) Establish the Project
   Office and structure, including Project Manager, Project Scientist,
   Science Advisory Committee and Legal and Environmental Issues Panel
   b) Define the scientific requirements. Terminates in
   Project Definition Review.
	\vspace{-0.1in}
 \item {\bf Conceptual Design (2010--2011)}: a) Perform trade studies on 
   cost, schedule, organization, performance and use; characterize
   the elements that will ensure delivery of the project requirements;
   Develop a Survey Operations Simulator; Select the final concept for
   the system identifying possible rescopes. 
   b) Produce a Conceptual Design Document, a System
   Requirements Document, a Management Plan, a System Engineering
   Plan, c) develop the Data Management architecture. Terminates in System Conceptual Design Review.
	\vspace{-0.1in}

\item {\bf Detailed Design (2012--2013)}: a) Perform detailed designs to meet
  System Requirements, b) Produce a Design Document, an Instrument Allocated
  Baseline with technical specifications, a complete 3D model and
  assembly diagrams, an Error Budget Document for mechanical
  parameters, and a Design Review Document. Terminates in Critical
  Design Review.
	\vspace{-0.1in}

 \item {\bf Project Construction (2014--2016)}: Construction of the
   SASIR camera, telescope, building, optics, control
   system. Terminates in Operation Review
	\vspace{-0.1in}

 \item {\bf SASIR Survey (2017--2021) and wrap-up activities
   (2022)}. SASIR Survey operations, including science meetings,
   postdoctoral positions, data archiving, data releases and outreach
   and extension activities.  Terminating in SASIR End Review
	\vspace{-0.1in}

\end{enumerate}

\noindent In the actual detailed project calendars, the above phases for
different subsystems or activities overlap in an out-of-phase
fashion. In particular, the project has decided not to develop a
completely new telescope concept, but to start from the Magellan or
MMT\index{MMT} telescope concept, in order to concentrate efforts on new
challenges like the camera concept. In particular, the primary mirror
has already been secured and is in the process of manufacturing. A
more complete breakdown of the activities in these phases is given in
the costing chart (\S \ref{sec:cost}).

\noindent{\bf Data for US Scientists}: 
The SASIR collaboration expects to release SASIR survey data to the US and Mexican astronomy communities incrementally during the science operations (following the models of 2MASS, SDSS, UKIDSS\index{2MASS}\index{SDSS}), with no more than a 18--20 month delay. Following the LSST model, transients will be released to the US and Mexico community at least as quickly as everyday, and possibly in near real-time.

\begin{figure}[h]  
        \centerline{\includegraphics[width=6.6in]{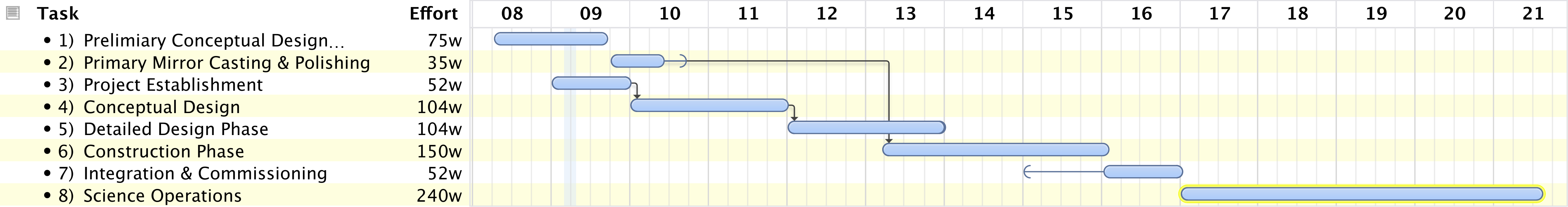}}
\caption{{\it Nominal SASIR schedule assuming ideal funding profile. See Figure \ref{fig:cost} for cost details.}}
\label{fig:gantt} 
\end{figure}

\subsection{Conceptual Design Phase Considerations}
\label{sec:cdc}
\vspace{-0.05in}

Upon securing funding for design work, we will initiate trade 
studies on cost, risk, and performance for various designs.  Early work will
concentrate on identifying and retiring the highest technical and cost risk areas. 
A second high-priority activity will be to clearly identify possible rescopes of the
current SASIR concept as part of technical/cost risk mitigation.
Table \ref{tab:study} summarizes some of the key issues to be analyzed
during the definition phase of the SASIR telescope.

\noindent{\bf Survey Operations Simulations}: In order to assess both
telescope design decisions and potential observing cadences on SASIR's
ability to deliver the science, a SASIR Operations Simulator will be
developed. The Simulator will accurately reflect telescope performance
and capabilities modeling all moving parts. It will utilize available
data to model weather and seeing. Finally, it will be able to
incorporate a variety of science programs with different goals and
cadences.  The results of the simulation will be put in a database for
ease of postprocessing.  Extensive multi-year simulations will allow
us to assess the potential output assuming a variety of design
decisions and optimize scientific programs so that the multiple
program goals can be attained over the course of the survey.
\vspace{-0.3in}
\begin{deluxetable}{lcl}
\tablewidth{7in}
\tabletypesize{\footnotesize}
\tablecaption{SASIR Telescope Key Parameters to be Studied \label{tab:study}}
\tablehead{
\colhead{ Aspect} &
\colhead{ Range to be Study} &
\colhead{ Main Questions/Issues to resolve}} 
\startdata
Field of View (FoV) \glossary{name={FOV},description={Field of View}} & 0.5$^\circ$--1.5$^\circ$ & 
  Science impact (survey speed).  \\
    && Cost and feasibility implications to M1/M2, \\
    && camera optics \& detectors.\\
Filter Set & YJHK -- JHK -- YJH & Science impact vs.\ technical
  feasibility and costs.\\
Survey Length & 3--9 yrs & Science-Impact/Running-Cost ratio.\\
Telescope Concept & 2-mirror, 3-mirror & Multi-band capability. K-emissivity. \\
   &w/wo re-imaging &  FoV. Optics complexity, size and feasibility. \\  
    &&  Scalability of known reference systems\\
    &&  (risk/critical-path mitigation) \\ 
    && Impact to base structural design (Magellan).\\
Multiple-Filter Strategy & Beam/Field splitting &  
   Science impact (band simultaneity). FoV. \\
    && Performance, cost and feasibility of overall design.\\
Pixel Size & 10--20 ${\mu}$m & Design merits versus camera complexity
    and cost. \\
    &&  Detector development status and costs.\\
Detector Scale & 0.2--0.4$''$/pixel & Optimization of image-quality budget
  and sampling, \\ 
    && sky background, system dimensions and performance.\\
Primary Conic Constant & 0.9--1.1 & M1 figuring. FoV. Image quality. \\
  & &   Subsystems simplification.\\
Focal reducer & With or without & Baffling complexity. Multi-band
  capabilities.\\
Collimated Stage & Partial \& Complete  &
  K-emissivity (white pupil). Multi-band \\
      & Refractive \& Reflective  & 
           capabilities (dichroic requirements).  \\
& optics & System complexity, \\
      & Uncollimated systems  & performance \&
          feasibility. Classic/Gregorian M2  \\
      && relation to collimator simplicity.\\
Focal Station(s) & Cassegrain, Nasmyth &
  Weight \& envelope limitations.
  Multiband capabilities.  \\
           & & FoV.\\
   & Prime foci & Impact and potential modifications of Magellan design.\\
Atmospheric Dispersion Corrector & location & Necessity for science programs, weight limitations
\enddata
\vspace{-0.2in}
\end{deluxetable}

\newpage
\section{Cost Estimates}
\label{sec:cost}
\vspace{-0.2in}
Aside from having decided on a  6.5 meter honeycomb primary
aperture, whose casting process has been initiated at the Arizona
Mirror Lab, and a concept for telescope structure and building based
on the Magellan/MMT\index{MMT}\index{Magellan} facilities, the SASIR design is at a preconception
level.  As SASIR is at the early phase of establishing the project, no
professional costing activity has been conducted.  As hereafter
detailed, the present SASIR budget estimate of \$170 Million
(FY2009, without contingency nor after-survey operations or
instrumentation), running from 2009 to 2022, is partially based on
the actual cost of the Magellan telescopes and complemented with our
best knowledge of present-day costs for the required subsystems
together with the expected management and qualified labor
needed. The primary mirror itself has been secured through a
partnership between INAOE\index{INAOE} and the University of Arizona and its cost
is already covered.

\begin{figure}[h]  
        \centerline{\includegraphics[width=6.3in]{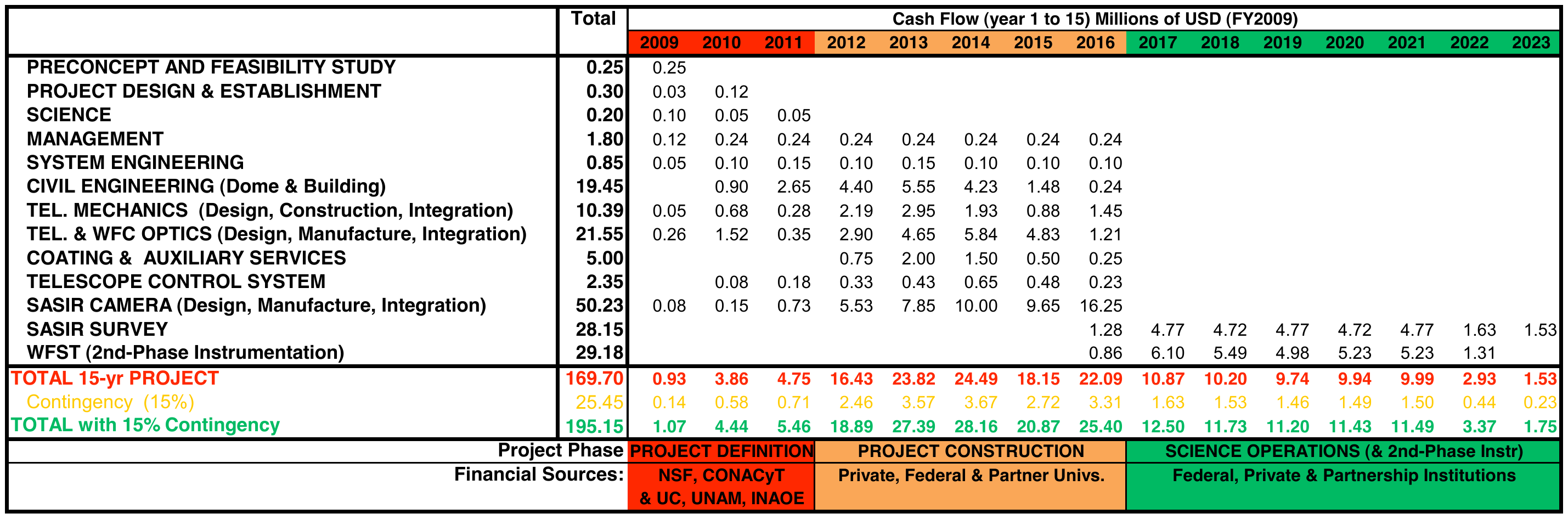}}
\caption{\small {\it Nominal costing cash flow organized in work packages.} }
\label{fig:cost} 
\end{figure}

\leftline{\bf Camera:} The cost and lead time of the detectors and the
readout electronics are well understood, with nominal costs (without
contingencies) for 124 + 10 spare arrays of $<$\$35M. This price makes
use of the ``buy the distribution'' notion, where, for instance, lower
quality (e.g. higher dark current) arrays from a foundry run would be
used in the $K$-band channel (which has a larger contribution from the sky).  Price per pixel is expected to drop by
more than a factor of 2 if the 4k x 4k lines become well established in the next few years.

With the lack of a more advanced design for this very important
component, the cost of the NIR SASIR Camera \index{SASIR!Camera}  is only indicative, and we
consider its costing as the least reliable estimate in the present
budget. Although the LSST\index{LSST} camera is not a strictly applicable reference,
nevertheless there are several reasons to expect the SASIR camera to
be less costly: 1) there are no moving parts inside the camera; 2) the
FoV is significantly smaller with a smaller aperture telescope; 3) the
SASIR detector planes are each three times smaller; 4) the field
corrector is not part of the camera and has much smaller optics.
Relative to other NIR imaging projects (e.g., VISTA, UKIDSS), \index{VISTA}\index{UKIDSS}\index{ODI} 
\glossary{name={ODI},description={One-Degree Imager}}
SASIR
is unique in considering four independent focal planes, avoiding the
need for filter-change mechanisms and permitting optics and detector
cost-saving optimizations with wavelength. In particular, SASIR detectors will likely be fabricated to be insensitive beyond 1.8 $\mu$m for the $Y, J,$ and $H$
arms, therefore requiring costly thermal-background blocking and a
cold-pupil solely in one arm ($K$).
\begin{deluxetable}{llc}
\tablewidth{6.8in}
\tablecaption{SASIR Costing Organized by Phases [``Federal'' component is assumed to be an equal split between US and Mexican national funding agencies] \label{tab:cost}}
\tablehead{\colhead{~} & \colhead{~} & \colhead{Nominal US National Share}}
\startdata
\hline
{\bf PROJECT ESTABLISHMENT (1yr)}:                                   & \centering \bf 0.77 & 0.19 \\
   Federal (50\%) and Institutional (50\%) Funds                    & ~ \\ 
\hline
\ \ \ \ SASIR Science Case \& Survey Definition                      & \centering 0.20 & ~ \\
\ \ \ \ Project Office, Manager \& Scientist                         & \centering 0.32 & ~ \\
\ \ \ \ Project Concept \& Feasibility Study                         & \centering 0.25 & ~ \\
\hline
{\bf CONCEPTUAL DESIGN (2yr):}                                         & \centering \bf 3.26 & 1.47 \\ 
   Federal (90\%) and Institutional (10\%) Funds                    & ~ \\
\hline
\ \ \ \ Project/Consortium Design \& Establishment                  & \centering 0.30 & ~ \\
\ \ \ \ Telescope Structure \& Dome PD \glossary{name={PD},description={Preliminary Design}}                              & \centering 0.35 & ~ \\
\ \ \ \ Wide-field Telescope Optical PD                              & \centering 0.30 & ~ \\
\ \ \ \ SASIR NIR Camera PD                                          & \centering 0.35 & ~ \\
\ \ \ \ Survey \& Operations PD                                      & \centering 0.03 & ~ \\
\ \ \ \ PDR                                                          & \centering 0.05 & ~ \\
\ \ \ \ Management \& PO Labor                                    & \centering 1.88 & ~ \\
\hline
{\bf DETAILED DESIGN (2yr): }                                          & \centering \bf 3.37 &  1.52\\
   Federal (90\%), Institutional (5\%) and Private (5\%) Funds      & ~ \\ 
\hline
\ \ \ \ Telescope Structure, Dome \& Mechanisms DD \glossary{name={DD},description={Detailed Design}}                  & \centering 0.30 & ~ \\ 
\ \ \ \ WF-telescope Global Configuration \& Optical DD              & \centering 0.54 & ~ \\ 
\ \ \ \ SASIR Camera DD                                              & \centering 0.50 & ~ \\
\ \ \ \ Dynamical and Kinematical models                             & \centering 0.05 & ~ \\
\ \ \ \ Survey \& Operations DD                                      & \centering 0.05 & ~ \\
\ \ \ \ CDR\glossary{name={CDR},description={Conceptual Design Review}}                                                          & \centering 0.05 & ~ \\
\ \ \ \ Management \& PO Labor                                    & \centering 1.88 & ~ \\
\hline
\bf CONSTRUCTION (3yr):                                              & \centering \bf 104.97 & 0 \\
   Federal (25\%), Private (70\%) and Institutional (5\%) Funds     & ~ \\
    (Mexico only)  & \\
\hline
\bf \ \ \ \ Civil Work                                               & \centering \bf 23.85 & ~ \\
\ \ \ \ \ \ Site Development and Main Infrastructures                & \raggedleft  7.50 & ~ \\
\ \ \ \ \ \ Building and Dome \index{SASIR!Telescope}                                       & \raggedleft  10.00 & ~ \\ 
\ \ \ \ \ \ Aux. Services: Cables/Hoses, Support Elements, Supplies  & \raggedleft  0.90 & ~ \\
\ \ \ \ \ \ Coating Facility \& Auxiliary Optical Services           & \raggedleft  5 & ~ \\
\ \ \ \ \ \ PO Labor (equiv. to 2 engineers over 3 years)         & \raggedleft  0.45 & ~ \\
\bf \ \ \ \ Telescope Structure \& Mechanism Construction                        & \centering \bf 9.05 & ~ \\
\ \ \ \ \ \ Structure and Mechanisms Manufacturing                   & \raggedleft  8.00 & ~ \\
\ \ \ \ \ \ Structure and Mechanisms Integration                     & \raggedleft  0.25 & ~ \\
\ \ \ \ \ \ Struc., Mechanisms \& Subsystems Acceptance Tests        & \raggedleft  0.10 & ~ \\
\ \ \ \ \ \ Structure \& Subsystems Alignment                        & \raggedleft  0.15 & ~ \\
\ \ \ \ \ \ Global Acceptance Tests                                  & \raggedleft  0.10 & ~ \\
\ \ \ \ \ \ PO Labor (equiv. to 2 engineers over 3 years)         & \raggedleft  0.45 & ~ \\
\bf \ \ \ \ Telescope \& Wide-Field Correcting Optics                           & \centering \bf 20.08 & ~ \\
\bf (Manufacture \& Integration) \\
\ \ \ \ \ \ Primary Mirror                                           & \raggedleft 10.00 & ~ \\ 
\ \ \ \ \ \ Secondary Mirror                                         & \raggedleft  2.50 & ~ \\
\ \ \ \ \ \ Calibration \& Guiding Services                          & \raggedleft  0.40 & ~ \\ 
\ \ \ \ \ \ Wide-Field Corrector \& other (TBC) optics               & \raggedleft  4.60 & ~ \\
\ \ \ \ \ \ Global Alignment                                         & \raggedleft  0.20 & ~ \\
\ \ \ \ \ \ PO Labor (equivalent to 2 engineers over 2.5 years)   & \raggedleft  0.38 & ~ \\
\bf \ \ \ \ Telescope Control System                                 & \centering \bf 2.35 & ~ \\
\ \ \ \ \ \ Hardware and Software                                    & \raggedleft  1.00 & ~ \\
\ \ \ \ \ \ PO Labor (equivalent to 6 engineers over 3 years)     & \raggedleft  1.35 & ~ \\
\bf \ \ \ \ SASIR Camera Manufacture \& Integration\index{SASIR!Camera}                                            & \centering \bf 48.63 & ~ \\
\ \ \ \ \ \ Optical components                                       & \raggedleft 10.50 & ~ \\
\ \ \ \ \ \ Structure. Mechanisms \& their Control                   & \raggedleft  1.00 & ~ \\
\ \ \ \ \ \ Cryogenics \& Control                                    & \raggedleft  1.50 & ~ \\
\ \ \ \ \ \ Detectors \& Control                                     & \raggedleft 33.00 & ~ \\
\ \ \ \ \ \ Camera Integration                                       & \raggedleft  0.20 & ~ \\
\ \ \ \ \ \ Camera/Telescope Integration                             & \raggedleft  0.05 & ~ \\
\ \ \ \ \ \ Pipelines and Survey Software                            & \raggedleft  2.00 & ~ \\
\ \ \ \ \ \ PO Labor (equivalent to 2 engineers over 2.5 years)   & \raggedleft  0.38 & ~ \\
\bf \ \ \ \ Project Management \& System Engineering                 & \centering \bf 1.02 & ~ \\
~ & ~ & ~ \\
\hline
\bf SASIR SURVEY (4--5yr):                                            & \centering \bf 28.15 & 12.67 \\
   Federal (90\%), Private (5\%) and Institutional (5\%) funds       & ~ \\
\hline
\ \ \ \ On site operations (5 yrs)   \index{SASIR!Survey}                                  & \centering 15.00 & ~ \\
\ \ \ \ Data Transfer, Processing, Management \& Storage             & \centering  3.70 & ~ \\
\ \ \ \ Post-Doctoral Positions (5 during 8 years)                   & \centering  3.00 & ~ \\
\ \ \ \ Postgraduate Fellowships (10 during 6 yrs)                   & \centering  3.20 & ~ \\
\ \ \ \ Science Workshops                                            & \centering  0.20 & ~ \\
\ \ \ \ Extension Activities                                         & \centering  2.00 & ~ \\
\ \ \ \ PO Labor (equiv.\ to 2 engineers over 7 years)             & \centering  1.05 & ~ \\
\hline
{\bf 2nd-Phase INSTRUMENTATION} &   TBD \\ 
\hline
\bf 2nd-Phase OPERATIONS      &   TBD \\ 
\hline
\bf PROJECT DISMANTLING        &  TBD \\ 
\hline
\enddata
\end{deluxetable}

\newpage
\begin{footnotesize}
	\begin{spacing}{0.93}
	\printglossary
	\end{spacing}
\end{footnotesize}
\newpage
%\twocolumn
\begin{spacing}{1.3}
%\bibliographystyle{hapj}
%\bibliography{journals_apj,decadal}

\begin{thebibliography}{32}
\expandafter\ifx\csname natexlab\endcsname\relax\def\natexlab#1{#1}\fi

\bibitem[{Blain {et~al.}(2009)}]{bla09}
Blain, A., {et~al.} 2009, {{Astro2010 Whitpaper};
  \href{http://www8.nationalacademies.org/astro2010/DetailFileDisplay.aspx?id=%
228}{link}}

\bibitem[{{Bloom} {et~al.}(2006){Bloom}, {Starr}, {Blake}, {Skrutskie}, \&
  {Falco}}]{bsb+06}
{Bloom}, J.~S., {Starr}, D.~L., {Blake}, C.~H., {Skrutskie}, M.~F., \& {Falco},
  E.~E. 2006, in Astronomical Society of the Pacific Conference Series, Vol.
  351, Astronomical Data Analysis Software and Systems XV, ed. C.~{Gabriel},
  C.~{Arviset}, D.~{Ponz}, \& S.~{Enrique}, 751

\bibitem[{{Bono}(2003)}]{bono03}
{Bono}, G. 2003, in Astronomical Society of the Pacific Conference Series, Vol.
  291, Hubble's Science Legacy: Future Optical/Ultraviolet Astronomy from
  Space, ed. K.~R. {Sembach}, J.~C. {Blades}, G.~D. {Illingworth}, \& R.~C.
  {Kennicutt}, Jr., 45

\bibitem[{{Brodwin} {et~al.}(2006){Brodwin}, {Lilly}, {Porciani}, {McCracken},
  {Le F{\`e}vre}, {Foucaud}, {Crampton}, \& {Mellier}}]{2006ApJS..162...20B}
{Brodwin}, M., {Lilly}, S.~J., {Porciani}, C., {McCracken}, H.~J., {Le
  F{\`e}vre}, O., {Foucaud}, S., {Crampton}, D., \& {Mellier}, Y. 2006, \apjs,
  162, 20, arXiv:astro-ph/0310038

\bibitem[{{Burrows} {et~al.}(2003){Burrows}, {Sudarsky}, \&
  {Lunine}}]{2003ApJ...596..587B}
{Burrows}, A., {Sudarsky}, D., \& {Lunine}, J.~I. 2003, \apj, 596, 587,
  arXiv:astro-ph/0304226

\bibitem[{Carilli {et~al.}(2009)}]{car09}
Carilli, C.~L., {et~al.} 2009, {{Astro2010 Whitpaper};
  \href{http://www8.nationalacademies.org/astro2010/DetailFileDisplay.aspx?id=%
62}{link}}

\bibitem[{{Cruz-Gonzalez} {et~al.}(2003){Cruz-Gonzalez}, {Avila}, \&
  {Tapia}}]{cg03}
{Cruz-Gonzalez}, I., {Avila}, R., \& {Tapia}, M., eds. 2003, Revista Mexicana
  de Astronomia y Astrofisica, vol. 27, Vol.~19, {San Pedro Martir :
  astronomical site evaluation}

\bibitem[{{Duval} {et~al.}(2004){Duval}, {Irace}, {Mainzer}, \&
  {Wright}}]{2004SPIE.5487..101D}
{Duval}, V.~G., {Irace}, W.~R., {Mainzer}, A.~K., \& {Wright}, E.~L. 2004, in
  Society of Photo-Optical Instrumentation Engineers (SPIE) Conference Series,
  Vol. 5487, Society of Photo-Optical Instrumentation Engineers (SPIE)
  Conference Series, ed. J.~C. {Mather}, 101--111

\bibitem[{{Fan} {et~al.}(2004){Fan}, {Hennawi}, {Richards}, {Strauss},
  {Schneider}, {Donley}, {Young}, {Annis}, {Lin}, {Lampeitl}, {Lupton}, {Gunn},
  {Knapp}, {Brandt}, {Anderson}, {Bahcall}, {Brinkmann}, {Brunner}, {Fukugita},
  {Szalay}, {Szokoly}, \& {York}}]{2004AJ....128..515F}
{Fan}, X. {et~al.} 2004, \aj, 128, 515, arXiv:astro-ph/0405138

\bibitem[{{Fata} {et~al.}(2004){Fata}, {Kradinov}, \& {Fabricant}}]{Fata04}
{Fata}, R.~G., {Kradinov}, V., \& {Fabricant}, D. 2004, in Society of
  Photo-Optical Instrumentation Engineers (SPIE) Conference Series, ed.
  A.~F.~M. {Moorwood} \& M.~{Iye}, Vol. 5492, 553--563

\bibitem[{{Faucher-Gigu{\`e}re} {et~al.}(2008){Faucher-Gigu{\`e}re}, {Lidz},
  {Hernquist}, \& {Zaldarriaga}}]{2008ApJ...688...85F}
{Faucher-Gigu{\`e}re}, C.-A., {Lidz}, A., {Hernquist}, L., \& {Zaldarriaga}, M.
  2008, \apj, 688, 85, 0807.4177

\bibitem[{{Feast}(2008)}]{feast08}
{Feast}, M.~W. 2008, 0806.3019, arXiv/0806.3019

\bibitem[{{Gonz{\'a}lez} \& {Orlov}(2007)}]{Gonz07}
{Gonz{\'a}lez}, J.~J., \& {Orlov}, V. 2007, in Revista Mexicana de Astronomia y
  Astrofisica, vol. 27, Vol.~28, Revista Mexicana de Astronomia y Astrofisica
  Conference Series, ed. S.~{Kurtz}, 60--66

\bibitem[{{Haiman} {et~al.}(2008){Haiman}, {Kocsis}, \&
  {Menou}}]{2008arXiv0807.4697H}
{Haiman}, Z., {Kocsis}, B., \& {Menou}, K. 2008, 0807.4697, arXiv/0807.4697

\bibitem[{{Kalas} {et~al.}(2008){Kalas}, {Graham}, {Chiang}, {Fitzgerald},
  {Clampin}, {Kite}, {Stapelfeldt}, {Marois}, \& {Krist}}]{2008Sci...322.1345K}
{Kalas}, P. {et~al.} 2008, Science, 322, 1345, arXiv/0811.1994

\bibitem[{{Kocsis} {et~al.}(2008){Kocsis}, {Haiman}, \& {Menou}}]{kocsis08}
{Kocsis}, B., {Haiman}, Z., \& {Menou}, K. 2008, \apj, 684, 870,
  arXiv/0712.1144

\bibitem[{{Lawrence} {et~al.}(2007){Lawrence}, {Warren}, {Almaini}, {Edge},
  {Hambly}, {Jameson}, {Lucas}, {Casali}, {Adamson}, {Dye}, {Emerson},
  {Foucaud}, {Hewett}, {Hirst}, {Hodgkin}, {Irwin}, {Lodieu}, {McMahon},
  {Simpson}, {Smail}, {Mortlock}, \& {Folger}}]{2007MNRAS.379.1599L}
{Lawrence}, A. {et~al.} 2007, \mnras, 379, 1599, arXiv:astro-ph/0604426

\bibitem[{{Madau} {et~al.}(2004){Madau}, {Rees}, {Volonteri}, {Haardt}, \&
  {Oh}}]{2004ApJ...604..484M}
{Madau}, P., {Rees}, M.~J., {Volonteri}, M., {Haardt}, F., \& {Oh}, S.~P. 2004,
  \apj, 604, 484, arXiv:astro-ph/0310223

\bibitem[{{McQuinn} {et~al.}(2009){McQuinn}, {Bloom}, {Grindlay}, {Band},
  {Barthelmy}, {Berger}, {Corsi}, {Covino}, {Fishman}, {Furlanetto}, {Gehrels},
  {Hartmann}, {Kouveliotou}, {Kutyrev}, {Loeb}, {Moseley}, {Piran}, {Piro},
  {Prochaska}, {Salvaterra}, {Schady}, {Soderberg}, \&
  {Tagliaferri}}]{2009arXiv0902.3442M}
{McQuinn}, M. {et~al.} 2009, ArXiv e-prints, arXiv/0902.3442

\bibitem[{{Michel} {et~al.}(2003){Michel}, {Hiriart}, \& {Chapela}}]{michel03}
{Michel}, R., {Hiriart}, D., \& {Chapela}, A. 2003, in Revista Mexicana de
  Astronomia y Astrofisica, vol. 27, Vol.~19, Revista Mexicana de Astronomia y
  Astrofisica Conference Series, ed. I.~{Cruz-Gonzalez}, R.~{Avila}, \&
  M.~{Tapia}, 99--102

\bibitem[{{Persson} {et~al.}(2008){Persson}, {Barkhouser}, {Birk}, {Hammond},
  {Harding}, {Koch}, {Marshall}, {McCarthy}, {Murphy}, {Orndorff},
  {Scharfstein}, {Shectman}, {Smee}, \& {Uomoto}}]{2008SPIE.7014E..95P}
{Persson}, S.~E. {et~al.} 2008, in Society of Photo-Optical Instrumentation
  Engineers (SPIE) Conference Series, Vol. 7014

\bibitem[{Poznanski {et~al.}(2008)}]{poznanski08}
Poznanski, D., {et~al.} 2008, arXiv/0810.4923

\bibitem[{Prochaska {et~al.}(2009)Prochaska, Bloom, Furlanetto, Fan, Gnedin,
  Oh, O'Shea, \& Stern}]{pbf+09}
Prochaska, J.~X., Bloom, J.~S., Furlanetto, S.~R., Fan, X., Gnedin, N., Oh, P.,
  O'Shea, B., \& Stern, D. 2009, {Astro2010 Whitpaper;
  \href{http://www8.nationalacademies.org/astro2010/DetailFileDisplay.aspx?id=%
253}{link}}

\bibitem[{{Schmidt} {et~al.}(1992){Schmidt}, {Kirshner}, \&
  {Eastman}}]{1992ApJ...395..366S}
{Schmidt}, B.~P., {Kirshner}, R.~P., \& {Eastman}, R.~G. 1992, \apj, 395, 366,
  arXiv:astro-ph/9204004

\bibitem[{{Schnittman} \& {Krolik}(2008)}]{2008ApJ...684..835S}
{Schnittman}, J.~D., \& {Krolik}, J.~H. 2008, \apj, 684, 835, arxiv/0802.3556

\bibitem[{{Schutz}(1986)}]{schutz86}
{Schutz}, B.~F. 1986, \nat, 323, 310

\bibitem[{{Skrutskie} {et~al.}(2006){Skrutskie}, {Cutri}, {Stiening},
  {Weinberg}, {Schneider}, {Carpenter}, {Beichman}, {Capps}, {Chester},
  {Elias}, {Huchra}, {Liebert}, {Lonsdale}, {Monet}, {Price}, {Seitzer},
  {Jarrett}, {Kirkpatrick}, {Gizis}, {Howard}, {Evans}, {Fowler}, {Fullmer},
  {Hurt}, {Light}, {Kopan}, {Marsh}, {McCallon}, {Tam}, {Van Dyk}, \&
  {Wheelock}}]{scs+06}
{Skrutskie}, M.~F. {et~al.} 2006, \aj, 131, 1163

\bibitem[{Sollima(2006)}]{scv06}
Sollima, C.~Cacciari, E.~V. 2006, MNRAS, 372, 1675

\bibitem[{{Tammann} {et~al.}(2008){Tammann}, {Sandage}, \&
  {Reindl}}]{tammann08}
{Tammann}, G.~A., {Sandage}, A., \& {Reindl}, B. 2008, \apj, 679, 52,
  arxiv/0712.2346

\bibitem[{{Tapia} {et~al.}(2007{\natexlab{a}}){Tapia}, {Cruz-Gonz{\'a}lez}, \&
  {Avila}}]{tca07}
{Tapia}, M., {Cruz-Gonz{\'a}lez}, I., \& {Avila}, R. 2007{\natexlab{a}}, in
  Revista Mexicana de Astronomia y Astrofisica, vol. 27, Vol.~28, Revista
  Mexicana de Astronomia y Astrofisica Conference Series, ed. S.~{Kurtz}, 9--15

\bibitem[{{Tapia} {et~al.}(2007{\natexlab{b}}){Tapia}, {Cruz-Gonz{\'a}lez},
  {Hiriart}, \& {Richer}}]{tap07}
{Tapia}, M., {Cruz-Gonz{\'a}lez}, I., {Hiriart}, D., \& {Richer}, M.
  2007{\natexlab{b}}, in Revista Mexicana de Astronomia y Astrofisica, vol. 27,
  Vol.~31, Revista Mexicana de Astronomia y Astrofisica Conference Series,
  47--60

\bibitem[{{Wood-Vasey} {et~al.}(2007){Wood-Vasey}, {Friedman}, {Bloom},
  {Hicken}, {Modjaz}, {Kirshner}, {Starr}, {Blake}, {Falco}, {Szentgyorgyi},
  {Challis}, {Blondin}, \& {Rest}}]{wood-vasey07b}
{Wood-Vasey}, W.~M. {et~al.} 2007, 0711.2068, arxiv/0711.2068

\end{thebibliography}

\end{spacing}

\newpage
\printindex

\end{document}